
\magnification = 1200
\def\lapp{\hbox{ {     \lower.40ex\hbox{$<$}
                   \atop \raise.20ex\hbox{$\sim$}
                   }     $}  }
\def\rapp{\hbox{$ {     \lower.40ex\hbox{$>$}
                   \atop \raise.20ex\hbox{$\sim$}
                   }     $}  }
\def\barre#1{{\not\mathrel #1}}

\def\krig#1{\vbox{\ialign{\hfil##\hfil\crcr
           $\raise0.3pt\hbox{$\scriptstyle \circ$}$\crcr\noalign
           {\kern-0.02pt\nointerlineskip}
$\displaystyle{#1}$\crcr}}}
\def\dbar#1{\vbox{\ialign{\hfil##\hfil\crcr
           $\raise0.3pt\hbox{$\scriptstyle =$}$\crcr\noalign
           {\kern-0.02pt\nointerlineskip}
$\displaystyle{#1}$\crcr}}}
\def\upar#1{\vbox{\ialign{\hfil##\hfil\crcr
          $\raise0.3pt\hbox{$\scriptstyle \leftrightarrow$}$\crcr\noali gn
          {\kern-0.02pt\nointerlineskip}
$\displaystyle{#1}$\crcr}}}
\def\ular#1{\vbox{\ialign{\hfil##\hfil\crcr
           $\raise0.3pt\hbox{$\scriptstyle \leftarrow$}$\crcr\noalign
           {\kern-0.02pt\nointerlineskip}
$\displaystyle{#1}$\crcr}}}

\def\svec#1{\skew{-2}\vec#1}
\def\Tr{\,{\rm Tr}\,}

\def\g5{\gamma_5}
\def\ks{\barre k}

\topskip=0.60truein
\leftskip=0.18truein
\vsize=8.8truein
\hsize=6.7truein
\tolerance 10000
\hfuzz=20pt

\baselineskip 12pt plus 1pt minus 1pt
\pageno=0
\centerline{\bf THRESHOLD PION ELECTROPRODUCTION}
\centerline{{\bf IN CHIRAL PERTURBATION THEORY}
 \footnote{*}{Work supported in part by Deutsche Forschungsgemeinschaft
under contract no. Me 864/2-2
and by Schweizerischer Nationalfonds.}}
\vskip 24pt
\centerline{V. Bernard$^1$, N. Kaiser$^2$, T. S.-H. Lee$^3$,
Ulf-G. Mei{\ss}ner$^{1,4,}$\footnote{$^\dagger$}{Heisenberg
Fellow.}}
\vskip 16pt
\centerline{$^1${\it Centre de Recherches Nucl\'{e}aires et Universit\'{e}
Louis Pasteur de Strasbourg}}
\centerline{\it Physique Th\'{e}orique,
BP 20Cr, 67037 Strasbourg Cedex 2, France}
\vskip  4pt
\centerline{$^2${\it Physik Department T30,
Technische Universit\"at M\"unchen}}
\centerline{\it
James
Franck Stra{\ss}e,
D-85747 Garching, Germany}
\vskip  4pt
\centerline{$^3${\it Physics Division,
Argonne National Laboratory}}
\centerline{\it Argonne,
Illinois 60439, USA}
\vskip  4pt
\centerline{$^4${\it Universit\"at Bern,
Institut f\"ur Theoretische Physik}}
\centerline{\it Sidlerstr. 5,
CH--3012 Bern,\ \ Switzerland}
\vskip 0.5truein
\baselineskip 12pt plus 1pt minus 1pt
\centerline{\bf ABSTRACT}
\baselineskip 12pt plus 1pt minus 1pt
\medskip
\noindent Electroproduction of pions on the nucleon near the threshold is
analyzed within the framework of baryon chiral perturbation theory.
We give a thorough discussion of
the low--energy theorems related to charged and neutral electropionproduction.
It is shown how the axial radius of the nucleon can be related to the
S--wave multipoles $E_{0+}^{(-)}$ and $L_{0+}^{(-)}$. The
chiral perturbation theory calculations of
the $\gamma^\star p \to \pi^0 p$ reaction
are found to be in good agreement with the recent near threshold data.
We also discuss the influence of some isospin--breaking
effects in this channel. For future experimental tests of the underlying
chiral dynamics, extensive predictions of differential cross sections and
multipole amplitudes are presented.
\vfill
\noindent BUTP--93/23 \hfill October 1993

\noindent CRN 93--45 \hfill
\eject
\baselineskip 16pt plus 1pt minus 1pt

\noindent{\bf I. INTRODUCTION AND SUMMARY}
\medskip
Threshold pion photo-- and electroproduction has received renewed
interest over the past few years from both the experimental and the
theoretical side. Extensive data were obtained for the process
$\gamma + p \to \pi^0 +p$. These data spurred numerous theoretical
investigations concerning the low--energy theorem for the electric
dipole amplitude $E_{0+}$ and also a critical reexamination of the
data was performed. In the framework of QCD these topics were addressed
in big detail in Ref.[1] (which from now on will be referred to as I).
Using virtual photons as probes one can get more detailed information
about the structure of the nucleon in the non--perturbative regime
of QCD due to the longitudinal coupling of the virtual photon to the
nucleon spin. For example, charged pion electroproduction encodes
information about fundamental quantities like e.g.
the axial form factor of the nucleon. Many of these topics are addressed
in the monograph by De Alfaro {\it et al.} [2].
Also, new experimental information
has recently become available for the process $\gamma^\star + p \to
\pi^0 +p$, where $\gamma^\star$ denotes the virtual photon [3].
This experiment is a major step beyond previous measurements
which were characterized by poor energy resolution and did not come
close enough to the production threshold. Therefore, the results of the
old measurements were dominated by the P--wave $M_{1+}$ multipole.
In contrast, the data presented in [3] were obtained in the energy
range of 0 to 2.5 MeV above threshold and at photon four--momenta
squared of $k^2 = -0.04$ to $-0.1$ GeV$^2$ ($k^2 < 0$ in the physical
region of the electroproduction reaction). More accurate data also
for the production of charged pions are expected to come from NIKHEF,
MAMI at Mainz and MIT Bates in the near future.

In this paper, we will be concerned with a systematic analysis of the
processes $\gamma^\star + p \to \pi^+ + n$,
$\gamma^\star + n \to \pi^- + p$ and
$\gamma^\star + p \to \pi^0 + p$ in the threshold region making use
of baryon chiral perturbation theory (CHPT). In general, CHPT allows
one to systematically investigate the strictures of the spontaneously
broken chiral symmetry of QCD. It is based on the observation that in
the three flavor sector of QCD, the quark masses are small and that
the theory in the limit of vanishing quark masses admits an exact
chiral symmetry. The latter is dynamically broken which leads to the
appearance of massless pseudoscalar excitations, the Goldstone bosons.
In the real world, the quark masses are not exactly zero and thus the
Goldstone bosons acquire a small mass. The interaction of these particles
with each other and matter fields like e.g. the nucleons are weak at
low energies as mandated by Goldstone's theorem. This fact is at the
heart of CHPT which amounts to a systematic and simultaneous
expansion of the QCD Green functions in small momenta and quark masses.
To perturbatively restore unitarity it is mandatory to consider pion
loop diagrams. In what follows, we will work in the one--loop
approximation which has been shown to be of sufficient accuracy for
many threshold phenomena. For a general review see Ref.[4]. At
next--to--leading order, one has not only pion loop graphs but also
local contact terms. The latter are accompanied by a priori unknown
coefficients, the so--called low energy constants. These can either
be determined from phenomenology [5] or estimated from resonance
exchange [6]. For the case under consideration, $\Delta(1232)$ and
vector meson exchanges,
the knowledge of some nucleon
electromagnetic radii as well as the pion charge radius allow us to
pin down all low energy constants appearing here.

The starting point of our investigation will be flavour SU(2) with
equal masses and we work to first order in the electromagnetic
coupling constant. As a consequence the one--loop approximation
does not include isospin--breaking
effects as they are revealed in the difference of the proton and
neutron or the charged and neutral pion masses. To minimally account
for these effects we will also present calculations in which the
respective particles have their physical masses. For the neutral
pion photoproduction it was argued in [7] that indeed the pion mass
difference is the most important effect of the isospin breaking.
In that reference it was also shown that this procedure leads to
a much improved description of the data for $E_{0+}^{\pi^0 p}$
in the threshold
region. We should stress, however, that ultimately a complete calculation
including effects of higher loops and of higher order in the fine
structure constant should be performed. This is beyond the scope of the
present paper.

The pertinent results of our investigation can be summarized as follows:
\medskip
\item{(i)} To one--loop order and to first order in the electromagnetic
coupling, there are 66 topologically different Feynman diagrams including pion
loops. These diagrams can be divided into three separately gauge invariant
subsets (the first class was already discussed in ref.[14]). After mass and
coupling constant renormalization, the corresponding invariant amplitudes
$A_i^{(a)}$ $(i=1,2,3,4,5,6;a=+,0,-)$ contain further UV divergences which can
be absorbed by two   low--energy constants related to the isovector charge
radius of the nucleon and the electromagnetic radius of the pion. In
addition, there are six  finite contact
terms of order $q^3$. Four of the unknown coefficients were already determined
in our study of threshold pion photoproduction [1] and the others can be fixed
from the axial and the isoscalar charge radii of the nucleon.
\medskip
\item{(ii)} We have discussed in some detail the low--energy theorems (LETs)
for the electric dipole amplitude $E_{0+}^{(a)}$ and the longitudinal
amplitude $L_{0+}^{(a)}$ $(a = +,0,-)$. For neutral pion electroproduction,
virtual pions generate infrared singularities which modify the familiar form
of the energy expansion [2,17]. In QCD, the expansion for $E_{0+}^{(a)}$ and
$L_{0+}^{(a)}$ in powers of $\mu
= M_\pi /m$ (the ratio of the pion to the nucleon mass) and $\nu = k^2 /m^2$
(with $k^2$ the four--momentum of the virtual photon) is given in eq.(5.1).
For $\gamma^\star p \to \pi^0 p$ and $\gamma^\star n \to \pi^0 n$, these agree
with the result given in ref.[14]. For the $(-)$ amplitudes which are probed
in charged pion electroproduction, we have found a modification of the LET
due to Nambu et al. [18]. This was already discussed in ref.[19] where it was
shown that an additional pion loop effect
modifies the relation between
$dE_{0+} (k^2) / dk^2$ at $k^2 = 0$ and the nucleon axial radius
$<r_A^2>^{1/2}$ (already in the chiral limit).
In fact, this correction leads to a better agreement
between the
axial radius determined from (anti)neutrino scattering [13] and from
pion electroproduction [31]. For the planned high precision
charged electroproduction  experiments it is mandatory to include this
effect in the analysis of the data. We have also worked out the terms
of order $\mu^3$ and $\mu^3 \ln \mu^3$ from the one--loop graphs for
$L_{0+}^{\pi^0 p} (\mu , 0 )$ and shown that the expansion in $\mu$
converges slowly. This agrees with our findings for photoproduction
reported in ref.[1].
\medskip
\item{(iii)} We have confronted the chiral prediction with the recent
very accurate NIKHEF data for $\gamma^\star p \to \pi^0 p$ [3]. As
already shown in ref.[30], one--loop effects are necessary to
understand the $k^2$--dependence of the S--wave cross section for
photon four--momenta $|k^2| \le 0.1$ GeV$^2$. The theoretical
prediction for $L_{0+}$ at the photon point, $|L_{0+}|^2 = 0.2 \,
\mu b$ is
in fair agreement with the result of ref.[3], $|L_{0+}|^2 = 0.13 \pm
0.05 \, \mu b$.
\medskip
\item{(iv)} The calculated multipole amplitudes are presented.
For future experimental tests of our predictions,
extensive results are also given for the transverse
($T$), longitudinal ($L$), interference ($I$) and
polarization ($P$) differential
cross sections $d\sigma_{T,L,I,P} / d\Omega$ and
are shown for all three channels, i.e.
$\gamma^\star p \to \pi^+ n$,
$\gamma^\star n \to \pi^- p$ and
$\gamma^\star p \to \pi^0 p$. To see the dynamical content of our CHPT
calculations, these results are
compared with the traditional calculations using pseudo-vector pion
coupling with the nucleon. Our results strongly suggest that to test the
dynamical content of the chiral perturbation theory prediction, it is
mandatory to perform a transverse/longitudinal separation of the data.
We have also shown that for photon four--momenta $|k^2| > 0.05$ GeV$^2$,
the loop corrections become large and thus it would be preferable
to have experiments at lower photon four--momenta.
\medskip
\item{(v)} For neutral electropionproduction of the proton, we have
investigated the dominant effect of isospin breaking, namely the
charged to neutral pion mass difference in the various loop functions.
We have given a gauge invariant prescription to implement this
effect. The resulting S--wave cross section is marginally different
from the one
presented in ref.[30].
Only at $k^2 = 0$, isospin breaking
seems to be of importance (due to the dominance of the class I
diagrams). We get an improved prediction
for $L_{0+}$ at the photon point, $|L_{0+}|^2 = 0.17 \, \mu b$.
At finite $k^2$, we could only find a substantial effect
on $d\sigma_P / d \Omega$ (at $k^2 = -0.04$ GeV$^2$). The largest
uncertainty therefore resides in the knowledge of the low--energy
constant $d_1$ (see ref.[1]).
\medskip

The paper is organized as follows. In section 2 we discuss some formal
aspects of pion electroproduction to fix our notation. The effective
Lagrangian which will be used is given in section 3. We only exhibit
the new terms related to the electroproduction reaction and refer the
reader to I for more detailed discussions. Section 4 contains
the invariant amplitudes for the reaction $\gamma^\star p \to \pi^0 p$.
Pion electroproduction low energy theorems are derived and discussed
in section 5. We also review critically previous attempts to formulate
these low energy theorems. The numerical results are presented in section
6 together with some proposals for future experiments. Various technicalities
are relegated to the appendices.
\bigskip
\noindent{\bf II. PION ELECTROPRODUCTION: SOME FORMAL ASPECTS}
\medskip
In this section, we define the current matrix elements for the
$\gamma^* N \rightarrow \pi N$ process in terms of
invariant amplitudes which will then be subject to the chiral expansion in
section IV. The formulas used in our calculations of
the $N(e,e^{\prime}\pi)N$ cross sections from
the current matrix elements can be found in Refs.[8,9] and are
therefore omitted
here. To see the dynamics at threshold, the multipole decomposition of the
current matrix elements at threshold is presented explicitly in this
section. The formulae needed for calculating the current matrix elements from
the multipole amplitudes tabulated in tables 1a,b,c
are given in Appendix C.
This will allow the
readers to calculate the CHPT predictions of
$(e,e^{\prime}\pi)$ spin observables accessible to possible future
experiments by using the well known formulae such as given
in Ref.[9].
The reader familiar with these topics is invited to skip
this section.
\medskip
\noindent{\bf II.1. BASIC CONSIDERATIONS}
\medskip
Consider the process $\gamma^*(k) + N(p_1) \to \pi^a(q) + N(p_2)$, with $N$
denoting a nucleon (proton or neutron), $\pi^a$ a pion with an
isospin index $a$ and
$\gamma^*$ the virtual photon with $k^2 < 0$. The case of photoproduction, $k^2
= 0$, was discussed in detail in I.
The pertinent Mandelstam variables
are $s=(p_1+k)^2, \,\, t=(p_1-p_2)^2$ and $u=(p_1-q)^2$ subject to the
constraint $s+t+u=2m^2+M_\pi^2 +k^2$. Here, $m$ and $M_\pi$ denote the nucleon
and pion mass, in order. To first order in the electromagnetic coupling, $e$,
to which we will work, the transition current matrix element is given as
$$\eqalign{
& J^\mu(p_2,s_2;q,a|p_1,s_1;k) = \cr
&i\bar u_2 \g5 \bigl\{ \gamma^\mu\ks B_1 + 2 P^\mu B_2 + 2 q^\mu  B_3 + 2 k^\mu
B_4 + \gamma^\mu B_5 + P^\mu \ks B_6 +k^\mu\ks B_7+q^\mu \ks B_8 \bigr\} u_1
\cr } \eqno(2.1)$$
with $s_j$ the spin index of nucleon $j$ and $P= (p_1+p_2)/2$. The amplitudes
$B_i(s,u)$ have the
conventional isospin decomposition (in the isospin symmetric limit),
$$B_i(s,u) = B_i^{(+)}(s,u)\,\, \delta_{a3} + B_i^{(-)}(s,u)\,\, {1\over 2}
[\tau_a, \tau_3] + B_i^{(0)}(s,u)\,\, \tau_a \,\,.\eqno(2.2)$$
Not all of the eight $B_i(s,u)$ are independent. Indeed, gauge invariance
$k_\mu J^\mu = 0$ requires
$$\eqalign{
& 2k^2 \,[B_1(s,u)+2B_4(s,u)] + (s-u)\,B_2(s,u) +2(s+u-2m^2)\,B_3(s,u) =0 \cr
& 4\,B_5(s,u) + (s-u)\,B_6(s,u) + 4k^2\,B_7(s,u)+ 2(s+u-2m^2) \, B_8(s,u)=0\cr
} \eqno(2.3)$$
so that one can express the transition current matrix element in terms of six
independent invariant functions, conventionally denoted by
$A_i(s,u),\,\,(i=1,...,6)$,
$$J^\mu = i \bar u_2 \g5 \sum_{i=1}^6 {\cal M}^\mu_i \, A_i(s,u)\, u_1
\eqno(2.4)$$
with
$$\eqalign{ &
{\cal M}_1^\mu = {1\over 2} (\gamma^\mu \ks - \ks \gamma^\mu), \qquad {\cal
M}_2^\mu = P^\mu ( 2q\cdot k - k^2) - P\cdot k (2 q^\mu - k^\mu),
\cr
& {\cal M}_3^\mu = \gamma^\mu q\cdot k - \ks  q^\mu , \qquad {\cal M}_4^\mu = 2
\gamma^\mu P\cdot k - 2 \ks P^\mu - m \gamma^\mu \ks + m\ks \gamma^\mu, \cr
& {\cal M}_5^\mu =  k^\mu  q \cdot k - q^\mu \, k^2, \qquad
{\cal M}_6^\mu = k^\mu\, \ks - \gamma^\mu \, k^2 \cr }\,. \eqno(2.5)$$
The $A_i(s,u) $ are related to the $B_i(s,u) $ via
$$\eqalign{ & A_1 = B_1 - m B_6, \qquad A_2 = {2B_2 \over M_\pi^2 - t}, \cr
& A_3 = - B_8, \qquad A_4 = -{1\over 2} B_6 , \qquad A_6 = B_7 , \cr
& A_5 = {2\over s + u - 2m^2} \biggl\{ B_1 + 2 B_4 + { (s-u)B_2 \over
2(t-M_\pi^2)} \biggr\} = {1\over k^2} \biggl\{ {s-u \over t -M_\pi^2} B_2 - 2
B_3 \biggr\} \,. \cr} \eqno(2.6)$$
For $A_5$ we have given two equivalent forms to make clear that the possible
zeros of $s+u -2m^2$ in the physical region do not lead to a pole (infinity) of
the current transition matrix element in the physical region.
In what follows, we will use both sets of invariant functions (see section IV).
Under
$(s \leftrightarrow u)$ crossing the amplitudes $A^{(+,0)}_{1,2,4},\,
A^{(-)}_{3,5,6},\,B^{(+,0)}_{1,2,6}$, $B^{(-)}_1+2B^{(-)}_4,\,
B^{(-)}_{3,5,7,8}$ are even, while $A^{(+,0)}_{3,5,6},\, A^{(-)}_{1,2,4},\,
B^{(+,0)}_1 +2B^{(+,0)}_4,\, B^{(+,0)}_{3,5,7,8},\,B^{(-)}_{1,2,6}$ are odd.

In terms of the isospin components, the physical channels under consideration
are
$$\eqalign{
J_\mu(\gamma^* p \to \pi^0 p ) & = J_\mu^{(0)} + J_\mu^{(+)} \cr
J_\mu(\gamma^* n \to \pi^0 n ) & = J_\mu^{(+)} - J_\mu^{(0)} \cr
J_\mu(\gamma^* p \to \pi^+ n ) & = \sqrt 2 [J_\mu^{(0)} + J_\mu^{(-)}] \cr
J_\mu(\gamma^* n \to \pi^- p ) & = \sqrt 2[ J_\mu^{(0)} - J_\mu^{(-)}]\,.
\cr}\eqno(2.7)$$
Having constructed the current transition matrix element $J_\mu$ it is then
straightforward to calculate observables. The pertinent kinematics and
definitions are outlined in refs.[8,9] to which the reader is referred for
details (see also Berends $et$ $al.$ [10]).
\medskip
\noindent{\bf II.2. MULTIPOLE DECOMPOSITION AT THRESHOLD}
\medskip
For the discussion of the low energy theorems in section V, we have to spell
out the corresponding multipole decomposition of the transition current matrix
element at threshold. In the $\gamma^* N$ center of mass system at threshold
$i.e.$ $q_\mu = (M_\pi, 0,0,0)$ one can express the current matrix element in
terms
of the two $S$-wave multipole amplitudes, called $E_{0+}$ and $L_{0+}$,
$$\vec J = 4\pi i(1+\mu)\, \chi_f^\dagger \bigl\{ E_{0+}(\mu,\nu) \,\vec \sigma
+ \bigl[L_{0+}(\mu,\nu) - E_{0+}(\mu,\nu)\bigr] \, \hat k \, \vec \sigma \cdot
\hat k  \bigr\} \chi_i \eqno(2.8)$$
with $\chi_{i,f}$ two component Pauli-spinors for the nucleon. For the later
discussion we have introduced the dimensionless quantities
$$\mu = {M_\pi \over m} , \qquad \nu = {k^2 \over m^2}\,. \eqno(2.9)$$
The multipole $E_{0+}$ characterizes the transverse and $L_{0+}$ the
longitudinal coupling of the virtual photon to the nucleon spin.
Alternatively to $L_{0+}$, we will also use the scalar multipole
$S_{0+}$ defined via:
$$S_{0+} (s, k^2 ) = {|\svec{k}| \over k_0} \, L_{0+} (s, k^2 )
\eqno(2.10)$$

At threshold, we can express $E_{0+}$ and $L_{0+}$ through the invariant
amplitudes $A_i(s,u)$ via (suppressing the isospin indices)
$$\eqalign{
E_{0+} & = {m\sqrt{(2+\mu)^2 - \nu} \over 8 \pi (1+\mu)^{3/2} }\biggl\{ \mu A_1
+ \mu m {\mu(2+\mu) + \nu \over 2(1+\mu)} A_3 + m {\mu(\mu^2 - \nu) \over
2(1+\mu)} A_4 - \nu m A_6\biggr\}\,, \cr
L_{0+} & = E_{0+} + {m\sqrt{(2+\mu)^2 - \nu} \over 16 \pi (1+\mu)^{5/2}}
(\mu^2 - \nu)\bigl\{ -A_1 - B_2 + B_1 + 2B_4 - \mu m A_4 - m(2+\mu) A_6\bigr\}
\cr }\eqno(2.11)$$
with the $A_i(s,u)$ and $B_i(s,u)$ evaluated at threshold $s_{th} =
m^2(1+\mu)^2$ and

\noindent
$u_{th} = m^2(1-\mu -\mu^2 + \mu \nu)/(1+\mu)$. This
completes the necessary formalism.

\bigskip
\noindent{\bf III. EFFECTIVE LAGRANGIAN}
\medskip
In this section, we will briefly discuss the effective lagrangian on which our
one-loop calculation in CHPT is based. We heavily borrow from our previous
work on threshold pion photoproduction with $k^2 = 0$. Here, we will only
discuss the terms which have to be added because of $k^2 <0$ in the
electroproduction case. For the details we refer the reader to I.

To systematically work out the consequences of the spontaneously broken chiral
symmetry at low energies, one makes use of an effective lagrangian of the
asymptotically
observed fields, in our case the Goldstone bosons (pions) and the
nucleons ($i.e.$ proton and neutron).
We work in flavor SU(2) and mostly in the isospin limit $m_u = m_d = \hat m$.
The proton and neutron are collected in the isodoublet Dirac-spinor,
$$\Psi= \left( \matrix {p \cr n\cr} \right) \eqno(3.1)$$
which transforms non-linearly under the chiral group $SU(2)_L\times SU(2)_R$,
$$\Psi(x) \to {\cal K}[g_L,g_R, U(x)] \, \Psi(x) \eqno(3.2)$$
where $g_{L,R}$ are group elements
of $SU(2)_{L,R}$ and the unitary unimodular
matrix $U(x)$ embodies the pion fields. For our calculation, it is most
convenient to work in the so-called $\sigma$-model gauge,
$$U(x) = [\sigma(x) + i \vec \tau \cdot  \vec{\pi}(x)
] /F, \qquad \sigma^2 + \svec
\pi^2 = F^2 \eqno(3.3)$$
with $F$ the pion decay constant in the chiral limit. To leading order, one
calculates tree diagrams from the effective lagrangain with the least number of
derivatives and quark mass insertions. It is given by
$${\cal L}^{(1)}_{\pi N} + {\cal L}_{\pi\pi}^{(2)} \eqno(3.4)$$
where the upper index indicates the so-called chiral power
according to the
counting rules
spelled out in I.
The explicit expressions for the lowest order
pion-nucleon, ${\cal L}^{(1)}_{\pi N} $, as well as the pion lagrangian, ${\cal
L}^{(2)}_{\pi\pi} $, are also given therein. At next-to-leading order, one has
two
types of contributions. These are the one pion loop diagrams which only involve
the few parameters occurring at lowest order. However, there are also local
contact terms which in general are necessary to renormalize the divergences of
the one-loop graphs. To one loop order, $i.e.$ chiral power $q^3$, their
generic form is
$${\cal L}^{(2)}_{\pi N} + {\cal L}^{(3)}_{\pi N} + \Delta {\cal L}^{(0)}_{\pi
N} + \Delta {\cal L}^{(1)}_{\pi N} + {\cal L}^{(4)}_{\pi \pi}\,.\eqno(3.5)$$
The terms $\Delta {\cal L}^{(0,1)}_{\pi N}$ relevant to renormalize the
nucleon mass and pion-nucleon coupling constant are discussed by GSS [11] and
in I.
In the meson sector, ${\cal L}_{\pi\pi}^{(4)}$ renormalizes the pion
decay constant $F_\pi$ and the pion mass and there
is one term which will give
extra contributions due to $k^2 <0$. It enters the pion charge form factor
[5],
$${\cal L}^{(4)}_{\pi\pi} = i { l_6 \over 2}
\Tr([ u^\mu, u^\nu] f^+_{\mu\nu})
\eqno(3.6)$$
with
$$\eqalign{
f_{\mu\nu}^\pm & = e (\partial_\mu A_\nu - \partial_\nu A_\mu) (uQu^\dagger \pm
u^\dagger Q u)\,,   \cr u^\mu & = i u^\dagger \nabla^\mu U u^\dagger ,
\qquad u =
\sqrt U \,.\cr } \eqno(3.7)$$
Here, $Q= {\rm diag}(1,0)$ is the charge matrix and $A_\mu$ denotes the
(external) photon field. The constant $l_6$ has a divergent piece (pole at
d=4), its finite part can be fixed from the
empirical value of the pion mean
square charge radius,
$$<r^2>_\pi  = - {1\over 8 \pi^2 F_\pi^2} \biggl( \ln{M_\pi \over \lambda} +
{1\over 2} \biggr) - {12\over F_\pi^2} l^r_6(\lambda) \eqno(3.8)$$
where the following renormalization prescription was used to cancel the
infinities from the loops
$$l_6 = - {L\over 6} + l^r_6(\lambda),\qquad L = {\lambda^{d-4} \over 16 \pi^2
} \biggl\{ {1\over d-4} +{1\over 2} (\gamma_E - 1 - \ln 4 \pi) \biggr\}
\eqno(3.9)$$
with $\lambda$ a scale introduced in dimensional regularization. Using the
empirical value

\noindent
$<r^2>_\pi = 0.439$ fm$^2$ [12], we have $l^r_6(1$ GeV) $ =
6.6\cdot 10^{-3}$. The other low-energy constants which occured already in
photoproduction are $l_3$ and $l_4$. These are related to the chiral
corrections of the  pion decay constant and pion mass at order $\hat m$ and
$\hat m^2$, respectively. In the $\pi N$ sector, there are three new terms
contributing to pion electroproduction at order $q^3$. These read
$$\eqalign{ {\cal L}^{(3)}_{\pi N} & = {b_9 \over F^2 } \bar \Psi \gamma^\mu
D^\nu f^+_{\mu\nu} \Psi + {\tilde b_9 \over F^2 } \bar \Psi \gamma^\mu \Psi \,
\Tr( D^\nu f^+_{\mu\nu}) \cr & + {g_A\over 12} b_{13} \, \bar \Psi
\gamma_5 \gamma^\mu \bigl( [ D^\nu , f^-_{\mu\nu} ] + {i \over 2}
[u^\nu
,f^+_{\mu\nu}]\bigr)  \Psi\,. \cr } \eqno(3.10)$$
The first two terms in eq.(3.10) are related to the electric mean square charge
radii of the proton and the neutron. It is well-known  that these radii develop
a logarithmic singularity in the chiral limit. The diagrams responsible for
this behaviour are in fact divergent and the constants $b_9, \,\tilde b_9$
absorb these infinities according to the renormalization prescription
$$b_9 = {L\over 6}(1-g_A^2) + b^r_9(\lambda), \qquad \tilde b_9 = {L\over 12}
(g_A^2- 1) + \tilde b^r_9(\lambda)\,. \eqno(3.11)$$
Notice  that the combination $b_9 + 2 \tilde b_9$ is finite. The pertinent
Dirac
form factors of the proton and neutron to one loop order read,
$$\eqalign{
F^p_1(k^2) = & 1 + {2k^2 \over F_\pi^2} [b^r_9(\lambda) + \tilde b^r_9(\lambda)
] + {k^2 (g_A^2 - 1) \over 96 \pi^2 F_\pi^2} \biggl( \ln{M_\pi \over \lambda} +
{1\over 2 }\biggr) + \biggl({g_{\pi N}^2 \over m^2} - {1\over F_\pi^2 } \biggl)
\dbar { J}_{22}^{\pi\pi}(k^2) \cr & + g_{\pi N}^2 \bigl\{ - 4 \bar {\tilde
\gamma}_3(k^2) - 16 m^2 \bar {\tilde \gamma}_4(k^2) - 2 \bar {\tilde
\Gamma}_3(k^2) + 4m^2 \bar {\tilde \Gamma}_4(k^2) + k^2[\tilde \Gamma_4(k^2)-
\tilde \Gamma_5(k^2)] \bigr\}\,, \cr
F^n_1(k^2) = &  {2k^2 \over F_\pi^2} \tilde b^r_9(\lambda)
 + {k^2 (1-g_A^2 ) \over 96 \pi^2 F_\pi^2} \biggl( \ln {M_\pi \over \lambda} +
{1\over2 }\biggr) + \biggl( {1\over F_\pi^2 }- {g_{\pi N}^2 \over m^2}  \biggl)
\dbar { J}_{22}^{\pi\pi}(k^2) \cr & + g_{\pi N}^2 \bigl\{     4 \bar {\tilde
\gamma}_3(k^2) + 16 m^2 \bar {\tilde \gamma}_4(k^2) - 4 \bar {\tilde
\Gamma}_3(k^2) + 8m^2 \bar {\tilde \Gamma}_4(k^2) + 2k^2[\tilde \Gamma_4(k^2)-
\tilde \Gamma_5(k^2)] \bigr\}\,. \cr } \eqno(3.12)$$
where the loop functions $\tilde \gamma_i(t) $ and $\tilde \Gamma_i(t) $ are
defined in appendix B.
{}From the empirical charge mean square radii, $<r^2_E>_p = 0.74$ fm$^2$ and
$<r^2_E>_n = - 0.12 $ fm$^2$ we can fix the coefficients at $\lambda = 1$ GeV
as $b^r_9(1$GeV) = 3.85$\cdot 10^{-3}$ and $\tilde b^r_9(1$GeV) = 1.45$\cdot
10^{-3}$. The last term in eq.(3.10) is related to the slope of the axial form
factor of the nucleon, $G_A(k^2)$. To one-loop order it can be written as
$$G_A(t) = g_A\biggl\{ 1 + {t \over 6} b_{13} + g_{\pi N}^2 \bigl[ - 2
\bar {\tilde \Gamma}_3(t) - 4m^2 \bar {\tilde \Gamma}_4(t) + t\bigl(\tilde
\Gamma_4(t)- \tilde \Gamma_5(t)\bigr) \bigr] \biggr\} \eqno(3.13)$$
The one loop contribution to the axial mean square
radius is finite and
very small, therefore the low-energy constant $b_{13}$ amounts
for most of its value. Using the average empirical value from (anti)-neutrino
scattering experiments, $<r^2_A> = (0.416 \pm 0.02)$ fm$^2$ [13]
we find
$b_{13} = (10.05 \pm 0.62)$ GeV$^{-2}$. All other terms in ${\cal
L}^{(2,3)}_{\pi N}$ which are relevant and already appeared in the
photoproduction process are discussed in I.
\bigskip
\noindent{\bf IV. INVARIANT AMPLITUDES TO ONE-LOOP}
\medskip
In this section, we will be concerned with the chiral expansion of the
invariant amplitudes of pion electroproduction as introduced
in section II. To
keep this section short we will give only the expressions for the reaction
$\gamma^* p \to \pi^0 p $ which was already discussed in the letter [14]. In
appendix A, the amplitudes for $\gamma^* n \to \pi^0 n$ as well as the minus
amplitudes $A_i^{(-)}(s,u), B^{(-)}_i(s,u)$ can be found. Together with the
ones
displayed here, these allow to calculate the charged electroproduction channels
$\gamma^* p \to \pi^+ n$ and $\gamma^* n \to \pi^- p$ by using eqs.(2.7). To
keep the notation compact, we will give $A_{1,3,4,6}(s,u),\,\, B_2(s,u)$ and
$B_1(s,u) + 2B_4(s,u)$ or an independent subset of the $B_i(s,u),\, (i \ne
3,5)$. These are then converted into the $A_i(s,u)$ via eqs.(2.6).
\medskip
\noindent{\bf IV.1. GENERAL REMARKS}
\medskip
We are seeking the chiral expansion of the invariant amplitudes $A_i(s,u)$,
$i.e.$ the expansion in small external momenta and quark masses,
$$A_i = A_i^{tree} + A_i^{1-loop} + A_i^{ct}\,. \eqno(4.1)$$
Here, $A_i^{tree}$ is the contribution calculated from the lowest order
effective lagrangian eq.(3.4). $A_i^{1-loop}$ stems from the one pion loop
diagrams generated by the lowest order vertices and involves only the
parameters $g_A, F$ and $m$ and finally $A_i^{ct}$ is calculated from tree
diagrams with exactly one insertion from ${\cal L}_{\pi\pi}^{(4)}$ or ${\cal
L}^{(2,3)}_{\pi N}$. They carry the information from the local counter terms
whose coefficients are not fixed by chiral symmetry requirements. Before giving
the explicit formulae, it is instructive to first discuss the general structure
of the $A_i(s,u)$ which emerges at one loop. As in the case of photoproduction
there are 66 topologically inequivalent one-loop diagrams. These can be grouped
into three separately gauge invariant classes, labelled class I, II and III,
respectively. While class I scales like $g_A \sim g_{\pi N}$ (if one uses the
Goldberger-Treiman relation), the diagrams in class II and III scale as $g_{\pi
N}^3$, cf. figs.2,3,4 in I,
with $g_{\pi N}$ the strong pion nucleon
coupling constant. In essence, any of the $A_i(s,u)$ has a contribution from
electric and magnetic Born terms, the non-pole loop contributions from classes
I,II and III plus the additional counter term contributions. The electric Born
term is the tree amplitude multiplied with the corresponding hadronic
electromagnetic form factor consistently calculated to one-loop and it involves
the physical values of the pion-nucleon coupling $g_{\pi N}$ and the nucleon
mass $m$. For the channel $\gamma^* p \to \pi^0 p$, the relevant form factor is
$F^p_1(k^2)$, the Dirac form factor of the proton. Similarly, the magnetic Born
term involves the proton Pauli form factor $F_2^p(k^2)$ as generated by some
graphs in class II and a counter term which allows to adjust the proton
anomalous magnetic moment. Notice that the latter is not present at tree level
but rather generated by loops and higher order counterterms. As in the
photoproduction case many of the one loop diagrams simply lead to a mass and
coupling constant renormalization (as discussed in I).
Another important topic concerns the isovector nucleon electric form
factor $F_1^v (k^2)$ and the pion electric form factor $F_\pi^v (k^2)$.
These appear naturally in some of the amplitudes exhibited below.
Since we are starting from a gauge invariant theory and because these
form factors are build up by loop and counterterm contributions in a
fashion that conserves the pertinent symmetries, there is no need of
setting $F_1^v (k^2) = F_\pi^v (k^2)$ as it is done in many model
calculations to preserve gauge invariance [2]. This feature is
particularly clearly shown in the discussion of the low--energy
theorems in section 5.
After these
general remarks, let us now discuss the various contributions to $A_i(s,u)$ for
the process $\gamma^* p \to \pi^0 p$ at one loop order.
\medskip
\noindent{\bf IV.2. ELECTRIC AND MAGNETIC BORN TERMS}
\medskip
The electric Born term modified by the proton Dirac form factor $F^p_1(k^2)$
(see eq.(3.12)) leads to the following amplitudes
$$\eqalign{
& A_1(s,u) = - B_2(s,u) = eg_{\pi N} \biggl( {1\over s - m^2} + {1\over u -
m^2} \biggr)\, F_1^p(k^2)\,, \cr
& B_1(s,u)+2B_4(s,u) = A_{3,4,6}(s,u) = 0 \,.\cr } \eqno(4.2)$$
It is important to notice that $F_1^p(k^2)$ is not the full physical form
factor but rather its one loop expansion. This is a point frequently overlooked
in the literature. We will come back to this point in section V in connection
with the discussion of the low energy theorems. A brief discussion of this
point has already been given in ref.[14]. Part of the magnetic Born term comes
from the counter term which corrects the proton anomalous magnetic moment,
$$\eqalign{
A_1(s,u) = & {e g_{\pi N} \over 2m^2} \, \delta \kappa_p , \qquad
A_{2,5,6}(s,u) = 0, \cr
A_3(s,u) = &- {eg_{\pi N}\over 2 m} \delta \kappa_p
\biggl( {1\over s - m^2} - {1\over u - m^2}
\biggr), \cr A_4(s,u) = & - {eg_{\pi N} \over 2 m} \delta \kappa_p
\biggl( {1\over s - m^2} +
 {1\over u - m^2} \biggr)\,. \cr } \eqno(4.3)$$
where
$$\delta \kappa_p = \kappa_p + 8 g_{\pi N}^2 m^2 \bigl\{ \Gamma_4(m^2,0)- 2
\gamma_4(m^2,0) \bigr\} = 1.288 \eqno(4.4)$$
is adjusted to give the empirical value of the proton anomalous magnetic
moment, $\kappa_p = 1.793$. In contrast to what was done in I,
we have
not subtracted the loop contribution to $F_2^p(k^2)$ from class II and
therefore only $\delta \kappa_p$ appears in eq.(4.3).
\medskip
\noindent{\bf VI.3. CONTRIBUTIONS FROM CLASSES I, II, III AND COUNTER TERMS}
\medskip
For class I and II the following representation is most economic
$$\eqalign{
A_1(s,u) & = a_1(s,k^2) + a_1(u,k^2) - B_2(s,u),\cr
B_2(s,u) & = {k^2 \over s - m^2} \bigl[ b_{14}(m^2,k^2) - b_{14}(s,k^2) \bigr]+
{k^2 \over u - m^2} \bigl[ b_{14}(m^2,k^2) - b_{14}(u,k^2) \bigr],
\cr B_1(s,u) + 2B_4(s,u) & = b_{14}(s,k^2) - b_{14}(u,k^2),\qquad
A_3(s,u)  = a_3(s,k^2) - a_3(u,k^2), \cr
A_4(s,u) & = a_3(s,k^2) + a_3(u,k^2),\qquad A_6(s,u) = a_6(s,k^2) -
a_6(u,k^2)\,.\cr} \eqno(4.5)$$

Class I, which scales as $g_{\pi N}$, leads to the following contributions,
$$\eqalign{
a_1(s,k^2) & = {2eg_{\pi N }\over F_\pi^2} (s-m^2) [ \gamma_6(s,k^2) -
\gamma_4(s,k^2) ], \cr
b_{14}(s,k^2) & = {eg_{\pi N} \over F_\pi^2} \bigl\{ (s+3m^2) [ 2
\gamma_6(s,k^2) - \gamma_2(s,k^2) ] +(s-m^2) [ \gamma_1(s,k^2) - 2
\gamma_5(s,k^2) ] \bigr\}, \cr
a_3(s,k^2) & = {4eg_{\pi N} \over F_\pi^2} m \gamma_4(s,k^2), \cr
a_6(s,k^2)  & = {2eg_{\pi N} \over F_\pi^2} m [ 2 \gamma_6(s,k^2) -
\gamma_2(s,k^2) ]\,.\cr } \eqno(4.6)$$
The various loop functions $\gamma_i(s,k^2)$ are defined in appendix B.

The class II diagrams give the following expressions proportional to $g_{\pi
N}^3$,
$$\eqalign{
a_1(s,k^2)  = {eg_{\pi N }^3\over 4m^2} & \bigl\{(s-m^2) [ 4\gamma_4(s,k^2) -
4\gamma_6(s,k^2)+ \Gamma_1(s,k^2) - \Gamma_2(s,k^2) \cr & -2 \Gamma_4(s,k^2) +
 2 \Gamma_6(s,k^2)  ]  + 8m^2 [2\gamma_4(s,k^2) - \Gamma_4(s,k^2) ]\bigr\},\cr
b_{14}(s,k^2)  = {eg_{\pi N}^3 \over 4m^2} & \bigl\{ (s+3m^2) [4\gamma_5(s,k^2)
- 2 \gamma_1(s,k^2)+ \Gamma_1(s,k^2) - \Gamma_2(s,k^2) - 2 \Gamma_5(s,k^2) ]
\cr & +{s^2 + 10 sm^2 + 5m^4 \over s-m^2}  [  2 \gamma_2(s,k^2) -
4\gamma_6(s,k^2)+ 2 \Gamma_6(s,k^2)  ] \bigr\} ,\cr
a_3(s,k^2)  = {eg_{\pi N}^3 \over 2m} & \bigl\{ 4\gamma_6(s,k^2) - 8
\gamma_4(s,k^2)+ 4\Gamma_4(s,k^2) - \Gamma_1(s,k^2) +  \Gamma_2(s,k^2) \cr & -
2 \Gamma_6(s,k^2)  + {8m^2 \over s - m^2} [ \Gamma_4(s,k^2) - 2
\gamma_4(s,k^2) ]\bigr\}, \cr
a_6(s,k^2)   = {eg_{\pi N}^3 \over 2m} & \bigl\{  4 \gamma_5(s,k^2)
-2\gamma_1(s,k^2) + \Gamma_1(s,k^2) - \Gamma_2(s,k^2)
-2\Gamma_5(s,k^2) \cr & + {4(s+m^2) \over s - m^2} [ \gamma_2(s,k^2) -
2\gamma_6(s,k^2) + \Gamma_6(s,k^2)    ] \bigr\}\,.\cr } \eqno(4.7)$$
Notice that the one loop representation contains the magnetic (Pauli) form
factor of the proton through the $k^2$ dependence of the loop functions. It is
proportional to the residue of $a_3(s,k^2) $ at $s= m^2$. Furthermore, one
observes that $a_6(s,k^2) $ and $b_{14}(s,k^2)$ do not have such poles since
$2\gamma_6(m^2,k^2) - \gamma_2(m^2,k^2) = 0 = \Gamma_6(m^2,k^2)$. This can be
easily seen from the Feynman representation eq.(B.2,3) by changing the
integration variable $x \to 1-x$.     Finally, we have to give the amplitudes
for class III, which are most tedious to work out. A compact representation is
obtained by giving the $B_i(s,u)$ $(i\ne 3,5)$,
$$\eqalign{B_1(s,u)  = &  eg_{\pi N}^3 \biggl\{
\Gamma_0^{\pi N}(s) -
{M_\pi^2 \over s - m^2} \underline \Gamma^{\pi N}_0(s) -
\Gamma_2^{\pi N}(s) +
(s-m^2) G_1(s,t,k^2) - 2 G_4(s,t,k^2) \cr & - k^2 G_5(s,t,k^2)
+(u-s) G_6(s,t,k^2) + (t+k^2 - M_\pi^2 ) G_7(s,t,k^2) -tG_9(s,t,k^2)
 \cr & +(t-4m^2) G_8(s,t,k^2) + 2 \Omega_4(s,u,k^2)  +
{s-m^2 \over 4m^2} \bigl[4\gamma_4(s,k^2)- 4\gamma_6(s,k^2) - \Gamma_1(s,k^2)
\cr & + \Gamma_2(s,k^2) + 2 \Gamma_4(s,k^2) - 2 \Gamma_6(s,k^2) \bigr]
+ {k^2 \over 4m^2} \bigl[- 2\gamma_1(s,k^2)  +
2 \gamma_2(s,k^2) + 4\gamma_5(s,k^2) \cr & - 4\gamma_6(s,k^2)  -\Gamma_1(s,k^2)
+  \Gamma_2(s,k^2) + 2 \Gamma_5(s,k^2) - 2 \Gamma_6(s,k^2)
\bigr]  \biggr\} + (s \leftrightarrow u)\,, \cr } $$
$$\eqalign{B_2(s,u) = & eg_{\pi N}^3 \biggl\{  -\Gamma_0^{\pi N}(s) +
{M_\pi^2 \over s - m^2} \underline \Gamma^{\pi N}_0(s) +  \Gamma_2^{\pi N}(s)
+ \tilde \Gamma_1(t) +
(s-m^2) \bigl[ G_2(s,t,k^2) \cr & -  G_3(s,t,k^2) + \Omega_2(s,u,k^2) - 2
\Omega_6(s,u,k^2) \bigr]   + 2G_4(s,t,k^2) + k^2 G_5(s,t,k^2) \cr & +
(2m^2-s-u) G_6(s,t,k^2)  + ( M_\pi^2 - t - k^2)
G_7(s,t,k^2)  +  t G_9(s,t,k^2) \cr & + (u-s-4m^2+M_\pi^2 - k^2) G_8(s,t,k^2)+
2(s-m^2 - M_\pi^2) \bigl[G_{10}(s,t,k^2) \cr & + \Omega_{10}(s,u,k^2) \bigr]
-2 \Omega_4(s,u,k^2) + 2(M_\pi^2 - 3 m^2 - s) \Omega_8(s,u,k^2) \cr & +
2\gamma_2(s,k^2)
+ {k^2 \over 4m^2} \bigl[ 2\gamma_1(s,k^2) - 2\gamma_2(s,k^2) -4\gamma_5(s,k^2)
+ 4\gamma_6(s,k^2) \cr & +\Gamma_1(s,k^2) - \Gamma_2(s,k^2) - 2 \Gamma_5(s,k^2)
+ 2 \Gamma_6(s,k^2) \bigr] \biggr\} + (s \leftrightarrow u)\,, \cr } $$

$$\eqalign{B_1(s,u) + & 2B_4(s,u) = eg_{\pi N}^3 \biggl\{
(s-m^2) \bigl[ G_1(s,t,k^2)+ G_2(s,t,k^2) - G_3(s,t,k^2) - 2 G_5(s,t,k^2)\cr &
+ \Omega_1(s,u,k^2) - \Omega_3(s,u,k^2) -2\Omega_5(s,u,k^2) \bigr] +
2( M_\pi^2 - 3 m^2 -s)\bigl[ G_6(s,t,k^2) \cr & - G_{10}(s,t,k^2)
+\Omega_6(s,u,k^2) - \Omega_{10}(s,u,k^2) \bigr]   + 2(2s-2m^2-M_\pi^2)
\bigl[ G_7(s,t,k^2) \cr & + \Omega_7(s,u,k^2) \bigr] + 2(m^2 + M_\pi^2 -s)
\bigl[ G_9(s,t,k^2) + \Omega_9(s,u,k^2) \bigr] + \gamma_0(s,k^2) \cr & -{s+3m^2
\over 2 m^2} \gamma_1(s,k^2)    + {s-m^2 \over 4 m^2} \bigl[ - \Gamma_1(s,k^2)
 + \Gamma_2(s,k^2) + 2\Gamma_5(s,k^2)  -  2
\Gamma_6(s,k^2) \cr & + 2 \gamma_2(s,k^2) +  4 \gamma_5(s,k^2) - 4
\gamma_6(s,k^2) \bigr] \biggr\} - (s \leftrightarrow u)\,, \cr } $$

$$\eqalign{ B_6(s,u) =  {e\over m}  g_{\pi N}^3 \bigl\{ & \Gamma_1(s,k^2) -
\Gamma_2(s,k^2) + 2 \Gamma_6(s,k^2) + 4\gamma_6(s,k^2) \cr & - 8m^2
G_8(s,t,k^2) - 8m^2 \Omega_8(s,u,k^2) \bigr\} + (s \leftrightarrow u)\,,\cr} $$

$$\eqalign{ B_7(s,u) =  {e\over 2 m}  g_{\pi N}^3 \bigl\{ & \Gamma_1(s,k^2) -
\Gamma_2(s,k^2) - 2 \Gamma_5(s,k^2) + 2\gamma_1(s,k^2) \cr & - 4\gamma_5(s,k^2)
 - 8m^2 G_6(s,t,k^2) + 8m^2 G_{10}(s,t,k^2) \cr & - 8m^2 \Omega_6(s,u,k^2) +
   8m^2 \Omega_{10}(s,u,k^2) \bigr\} - (s \leftrightarrow u)\,, \cr } $$

$$\eqalign{ B_8(s,u) =  {e\over 2m}  g_{\pi N}^3 \bigl\{ & \Gamma_1(s,k^2) -
\Gamma_2(s,k^2) + 2 \Gamma_6(s,k^2) + 4\gamma_6(s,k^2) \cr & - 8m^2
G_{10}(s,t,k^2) - 8m^2 \Omega_{10}(s,u,k^2) \bigr\} - (s \leftrightarrow u)\,.
\cr } \eqno(4.8)$$

An excellent numerical check on these rather involved expressions is given by
the gauge invariance relations eq.(2.3) (for that, one also calculates
$B_{3,5}(s,u)$), with the exception of $B_1(s,u)$ which is not constrained by
the gauge invariance condition $k_\mu  J^\mu = 0$. As a further check, all
the above expressions of course match to the formulae given in I
for the
case $k^2 = 0$. Finally, for the channel $\gamma^* p \to \pi^0 p $ there is one
counter term contribution to $A_4(s,u) = a_4^{\pi^0 p}$ from ${\cal L}_{\pi
N}^{(3)}$. The same term was already present in the photoproduction calculation
and its coefficient was estimated from resonance saturation. Equivalently, one
can say that the determination of the constant $a_4^{\pi^0 p}$ in I
amounted to an overall best fit to the total photoproduction cross section for
$\gamma p \to \pi^0 p $ in the threshold region based on the Mainz data [15]
(with $E_\gamma \le 160$ MeV).

With the invariant amplitudes given here and in appendix A, we are now in the
position of discussing the pertinent physics issues related to threshold pion
electroproduction.
\bigskip
\noindent{\bf V. LOW ENERGY THEOREMS}
\medskip
\noindent{\bf V.1. GENERAL REMARKS}
\medskip
Pion electroproduction low energy theorems (LETs) have been discussed in many
articles [16]. One of the
   most recent discussions is due to Scherer and Koch [17]. The
underlying idea is to expand the threshold $S$-wave multipoles $E_{0+}$ and
$L_{0+}$\footnote{*}{Here $E_{0+}$ and $L_{0+}$ stand as generic symbol for any
isospin channel.} in terms of the dimensionless and small parameters $\mu =
M_\pi/m\simeq 1/7$ and $\nu = k^2/m^2$. Evidently, $\nu $ can be made arbitrary
small as $k^2$ approaches zero, the so-called photon point. One of the main
differences to the photoproduction case was first pointed out by Nambu, Luri\'e
and Shrauner (NLS) [18], namely that charged electroproduction involves the
axial form factor of the nucleon $G_A(t)$. It therefore gives an other
possibility of determining this fundamental quantity. However, the LET of NLS
refers strictly to the chiral limit and one therefore
has to add corrections for
the physical case of a finite pion mass. Such corrections are discussed for
example in the monograph [2].
As we will show in some detail later, CHPT allows
to systematically calculate the next-to-leading order corrections to
the LET of NLS, as already reported in ref.[19]. This is only one particular
example of the chiral expansion done consistently. In fact, the one loop
calculation performed here allows us to calculate all terms of the
multipole amplitudes $E_{0+}(\mu,\nu)$ and $L_{0+}(\mu,\nu)$ up to and
including order ${\cal O}(\mu^2, \nu)$. The results of this calculation will be
reported below.

First, however, some claryfying words about the meaning of LETs and their
relation to the CHPT calculation are in order since in the present literature
one finds many erroneous statements. In QCD, the quark masses are   a priori
free parameters. For the two flavor sector, one can set $m_u = m_d = 0$ to a
good approximation. In this limit, the so-called chiral limit, the pions are
exactly massless, as mandated by the Goldstone theorem. CHPT now allows to
systematically work out the consequences of the spontaneous and the
explicit chiral symmetry
breaking in QCD, as an expansion in small external momenta and quark masses.
For the case of pion electroproduction at threshold, these small expansion
parameters are the pion mass $M_\pi^2 \sim \hat m$ and the photon four-momentum
$k^2$. As it is well-known and most clearly spelled out by Weinberg [20], CHPT
embodies the very general principles of gauge invariance, analyticity, crossing
and PCAC (pion pole dominance). Therefore, to lowest order, one recovers the
venerable current algebra LETs, which are based on these principles. The
effective chiral lagrangian is simply a tool to calculate these but also all
next-to-leading order corrections in a systematic and controlled fashion. The
question of interest here is now: What are the corrections to the lowest order
statements for $E_{0+}(\mu,\nu)$ and $L_{0+}(\mu,\nu)$ at threshold? This
question can be unambigosly answered by making use of the CHPT machinery. For
that it is mandatory to consider pion loop diagrams. It was already shown in
the photoproduction case that the conventional LET due to Vainsthein and
Zakharov [21] and de Baenst [22] was incomplete and has to be modified at
next-to-leading order, ${\cal O}(\mu^2)$, due to a logarithmic singularity of
certain loop diagrams in the chiral limit of vanishing pion mass. Such an
effect can never be found without explicit calculation of all relevant
diagrams. In the light of this result we also expect similar modifications to
show up in pion electroproduction.  This expectation was already borne out by
an explicit calculation of the class I diagrams for the reaction $\gamma^* p
\to \pi^0 p $ [14].
In the following section, we will give the complete list of LETs
for electropion production (in the isospin basis).

In view of the above arguments, it should be clear that simply calculating Born
(tree) diagrams supplemented by form factors (in a gauge invariant fashion) can
not give all corrections at next-to-leading order [17,23,24].
To be more specific,
consider first charged pion electroproduction. The leading term in
$E_{0+}(\mu,\nu)$ and $L_{0+}(\mu,\nu)$ are of order one, which is nothing but
the
generalization of the famous Kroll-Ruderman theorem [25].
In ref.[17] it
was claimed that all corrections up to and including ${\cal O}(\mu^2,\nu)$ can
be calculated from PCAC and current conservation without considering loop
diagrams. From the
above remarks on a consistent expansion in CHPT, it should be
obvious that this expectation is too naive. It is based on the incorrect
assumption that while PCAC is a very general principle, CHPT is derived from a
particular lagrangian and would be more restrictive. Quite contrary, any theory
or model
which claims to embody PCAC should lead to the same result as CHPT if one goes
to the same level of sophistication. Similarly, the recent discussion of
Ohta [24] is misleading. When calculating the LETs
in QCD, one never has
to worry about half off--shell nucleon form
factors and thus the findings reported in refs.[14,19] are model--independent
and do not rely on any assumptions about these form factors.
We hope that with these remarks we have
been able to shed some light on the seemingly controversal theoretical
interpretation of the LETs. In fact, the meaning of the LETs is unique
(experimental quantities are expanded in the chiral symmetry breaking parameter
$M_\pi$ and small external momenta) and their accuracy at a given order can be
tested experimentally. Let us now present the pertinent complete results at
next-to-leading order.
\medskip
\noindent{\bf V.2. FORMULATION AND DISCUSSION OF THE LOW ENERGY THEOREMS}
\medskip
The derivation of the LETs for $E_{0+}^{(+,0,-)}$ and $L_{0+}^{(+,0,-)}$ in
electropion production is a straightforward generalization of the
photoproduction case investigated in I.
Here, we only wish to point out
that for this calculation it is most convenient to make use of the heavy mass
formulation of baryon CHPT [26]. It was shown in ref.[27] that a one loop
calculation in this framework reproduces all relevant terms. We therefore skip
the computational details here. The correct form of the LETs for
$E_{0+}(\mu,\nu)$ and $L_{0+}(\mu,\nu)$ in the various isospin channels
$(+,0,-)$ at next-to-leading order is given by\footnote{$^*$}{Some of the
kinematical prefactors in eqs.(5.1) have not been expanded to keep the
notation compact.}
$$\eqalign{
E_{0+}^{(+)}(\mu,\nu)  = & {eg_{\pi N} \over 32 \pi m} \biggl\{ - 2 \mu + \mu^2
(3+\kappa_v) - \nu(1+\kappa_v) + {\mu^2 m^2 \over 4\pi F_\pi^2}\, \Xi_1(-\nu
\mu^{-2}) \biggr\} + {\cal O}(q^3)\,, \cr
L_{0+}^{(+)}(\mu,\nu)  = & E_{0+}^{(+)}(\mu,\nu) + {eg_{\pi N} \over 32 \pi
m}(\mu^2 - \nu) \biggl\{ - \kappa_v + { m^2 \over 4\pi F_\pi^2}\, \Xi_2(-\nu
\mu^{-2}) \biggr\} + {\cal O}(q^3)\,, \cr
E_{0+}^{(0)}(\mu,\nu)  = & {eg_{\pi N} \over 32 \pi m} \bigl\{ - 2 \mu + \mu^2
(3+\kappa_s) - \nu(1+\kappa_s)  \bigr\} + {\cal O}(q^3)\,, \cr
L_{0+}^{(0)}(\mu,\nu)  = & E_{0+}^{(0)}(\mu,\nu) + {eg_{\pi N} \over 32 \pi m}
(\nu - \mu^2) \, \kappa_s + {\cal O}(q^3)\,, \cr
E_{0+}^{(-)}(\mu,\nu)  = &  {eg_{\pi N} \over 8 \pi m} \biggl\{1 - \mu +C\mu^2
 +\nu \biggl( {\kappa_v \over 4} + {1\over 8} + {m^2 \over 6} <r^2_A> \biggr) +
 {\mu^2 m^2 \over 8\pi^2 F_\pi^2} \,\Xi_3(-\nu \mu^{-2}) \biggr\} +
{\cal O}(q^3)\,, \cr
L_{0+}^{(-)}(\mu,\nu)  = & E_{0+}^{(-)}(\mu,\nu) + {eg_{\pi N} \over 8 \pi m}
(\mu^2 - \nu)
\biggl\{ {\kappa_v \over 4} + {m^2 \over 6} <r^2_A> + {\sqrt{(2+\mu)^2 - \nu}
\over 2(1+\mu)^{3/2} (\nu - 2\mu^2 - \mu^3)}\cr & +
\biggl( {1\over \nu - 2\mu^2} - {1\over \nu} \biggr) \bigl( F_\pi^v(m^2 \nu) -
1 \bigr) +  { m^2 \over 8\pi^2 F_\pi^2}\, \Xi_4(-\nu \mu^{-2}) \biggr\} +
{\cal O}(q^3)\,. \cr}  \eqno(5.1)$$
where the functions $\Xi_j(- \nu/\mu^2), \,\,(j = 1,2,3,4)$ can not be further
expanded since the argument $-\nu/\mu^2$ counts as order one. They are given by
$$\eqalign{
\Xi_1(\rho)  = & {\rho \over 1+ \rho} + {(2+\rho)^2 \over 2(1+\rho)^{3/2}}
\arccos{-\rho \over 2+ \rho}\,, \cr
\Xi_2(\rho)  = & {2-\rho \over (1+ \rho)^2} - {\rho^2 +2\rho +4 \over
2(1+\rho)^{5/2}} \arccos{-\rho \over 2+ \rho} \,,\cr
\Xi_3(\rho)  = &  \sqrt{1+ {4\over \rho}} \ln \biggl( \sqrt{1+ {\rho \over
4}} + {\sqrt \rho \over 2} \biggr) \cr & + 2 \int_0^1 dx
\sqrt{(1-x)[1+x(1+\rho)] } \arctan {x\over \sqrt{(1-x)[1+x(1+\rho)]}}\,, \cr
\Xi_4(\rho)  = &\int_0^1 dx{x(1-2x) \over \sqrt{(1-x)[1+x(1+\rho)] }}
\arctan {x\over \sqrt{(1-x)[1+x(1+\rho)]}}\,. \cr}  \eqno(5.2)$$
These functions are shown in fig.[1] for $0\le \rho \le 10$. They exhibit a
very smooth behaviour. $\kappa_v = \kappa_p - \kappa_n = 3.71$ and $\kappa_s =
\kappa_p + \kappa_n = -0.12$ are the isovector and isoscalar anomalous magnetic
moment of the nucleon, respectively. We have to make various remarks about the
LETs exhibited in eqs.(5.1). First, adding the $(+)$ and $(0)$ multipoles, one
recovers the LETs for $\gamma^* p \to \pi^0 p$ derived in refs.[14,27].
One sees
that class II and III do not at all contribute at order ${\cal O}(\mu^2 ,
\nu)$. This feature is most easily understood in the heavy mass formulation of
CHPT, where simple selection rules determine those few diagrams which are
non-vanishing at threshold (see ref.[27]). Second, notice that the LETs do
not contain the full
electromagnetic form factors of the nucleon. This is due to the
fact that in the consistent chiral power counting on which CHPT is based these
form factor effects are already of order ${\cal O}(\mu   \nu )= {\cal O}(q^3)$
and therefore consistently have to be dropped. Furthermore,
one sees only the
normalization        of the magnetic form factors, $i.e.$ the respective
anomalous magnetic moments. Of particular interest are the LETs for the $(-)$
amplitudes. In the chiral limit, the LET for $E_{0+}^{(-)}$ agrees with the one
of NLS [18] at order ${\cal O}(\nu)$. The constant $C$ which appears also in
photoproduction is of order one and it is given by
$$C ={9\over 8} - {m^2 \over 8\pi^2 F_\pi^2} \bigl( 1 + \ln \mu\bigr) + {2m^3
\over e g_{\pi N} } \bigl\{ 4 ma_1^{(-)} + a_3^{(-)} \bigr\} \eqno(5.3)$$
where the coefficients $a_{1,3}^{(-)}$ have been estimated in ref.[1] using
resonance saturation. For the central values given there, $a_1^{(-)} = 1.4$
GeV$^{-4}$ and $a_3^{(-)} = -9.9$ GeV$^{-3}$, we find $C = 0.40$. In ref.[19],
the consequences of the LET for $E_{0+}^{(-)} $ were discussed. If one expands
$\Xi_3(-\nu\mu^{-2})$ in powers of $k^2 = \nu m^2 $ and picks up all terms
proportional to $\nu$ ($i.e.$ all terms which contribute to the slope of
$E_{0+}^{(-)}$ at $k^2 = 0$), one finds
$${\partial E_{0+}^{(-)} \over \partial k^2 } \bigg|_{k^2 = 0} = {2\kappa_v + 1
\over 8m^2} + {1\over 6} <r^2_A> + {1\over 128 F_\pi^2} \biggl( 1 - {12 \over
\pi^2} \biggr) + {\cal O}(\mu) \eqno(5.4)$$
where the first two terms on the right hand side of eq.(5.4) have first been
given by NLS. As discussed in ref.[19], the new term at order $k^2$ (not
vanishing in the chiral limit)
due to loop effects has a numerical value of $ -
0.0456$ fm$^2$ and this closes the gap between the experimental determination
of the nucleon axial mean square radius $<r^2_A>$ from neutrino experiments,
$<r^2_A>_\nu = 0.42$ fm$^2$, and pion electroproduction analysis,
$<r^2_A>_{\gamma^*} = 0.37$ fm$^2$ [31].
To get an idea about the higher orders in
$k^2$, we show in fig.[2] the two functions
$$\eqalign{
\Phi_1(k^2) & = 1 + {k^2 \over 6} <r^2_A>\,, \cr
\Phi_2(k^2) & = 1 + {k^2 \over 6} <r^2_A> + {M_\pi^2 \over 8 \pi^2 F_\pi^2 }
\biggl\{ \Xi_3\bigl(-{k^2 \over M_\pi^2} \bigr) -{\pi^2\over 8} - {1\over 2}
\biggr\} \cr } \eqno(5.5)$$
for $<r^2_A> = 0.42 $ fm$^2$. For $|k^2| \le 0.5$ GeV$^2$, the corrections of
order $k^4$ (and higher) from one loop
are quite small. Most important for the analysis of
new precise electroproduction experiments is that one takes into account the
novel term at order $k^2$ appearing in eq.(5.4). Let us now turn to the LET for
$L_{0+}^{(-)} - E_{0+}^{(-)}$ in eq.(5.1). In contrast to the electroweak
nucleon form factors, the full one loop expression of the pion charge form
factor $F_\pi^v(k^2) $ enters. This can be traced back to the following fact.
The one loop representation of $F_\pi^v(k^2)$ takes the form
$$F_\pi^v(k^2) - 1 = {1\over F_\pi^2 } \biggl\{ c_1
k^2 \ln M_\pi +  c_2 {k^4 \over
M_\pi^2} + c_3 {k^6 \over M_\pi^4} + \dots \biggr\} \eqno(5.6)$$
and all terms in the curly bracket are of the same order ${\cal O}(q^2)$. This
feature is similar to
the functions $\Xi_j(-\nu \mu^{-2})$ whose expansion can not
be truncated at any finite order. To the order we are working, one can even
identify the one loop pion charge form factor with the empirical one, since
the differences are order $q^4$. To get an idea of the higher order
corrections, we can use the two loop representation of ref.[28]. At $k^2 = -
0.2$ GeV$^2$, $i.e.\,\, \nu = -0.23$, one finds that
the one loop and the two loop representation
of $F_\pi^v(k^2) $ differ  by $22\%$. Finally, let us stress again the two
salient features of the LETs presented in eqs.(5.1). First, in some cases
virtual pion loops generate contributions with chiral singularities which
modify the form of the LETs based on an incomplete calculation of tree
diagrams including electroweak form factors. These unfamiliar terms are given
by the $\Xi$-functions. Second, in the consistent
expansion only the first moment of the nucleon electromagnetic form factors and
the first and  second moment of the nucleon axial form factor survive. The pion
charge form factor, however,  enters with its full one loop expression. The
LETs presented in eqs.(5.1) are the ones which follow from the spontaneous (and
explicit) beaking of chiral symmetry in QCD.

It might be instructive to investigate how fast the convergence of the LETs is.
In the photoproduction case it was shown that the ${\cal O}(\mu^3)$
contributions of the one loop approximation to $E_{0+}^{\pi^0 p}(\mu) $
substantially reduce the ${\cal O}(\mu^2)$ terms. Such a trend can also be seen
here.
To be more specific, consider $L_{0+}^{\pi^0 p}(\mu, 0)$ at the photon
point. For that we have evaluated all the class I contributions
up-to-and-including ${\cal O}(\mu^3, \mu^3 \ln \mu)$. The result reads
$$\eqalign{
L_{0+}^{\pi^0 p} (\mu , 0) & = {eg_{\pi N} \over8\pi m}\biggl\{ - \mu + {3\over
2} \mu^2 - {15\over 8} \mu^3 + {m^2 \mu^2 \over 8 \pi F_\pi^2} \biggl[ 1 +
{4\over \pi} \mu \ln \mu + \mu \biggl( {1\over \pi} - {3\over 2} - {\pi \over
8} \biggr) \biggr] \biggr\} \cr
& = -  0.0247 \cdot ( 1 - 0.797 + 0.377) \,\,{\rm GeV}^{-1} = - 0.0143 \,\,
{\rm GeV}^{-1} \cr }\eqno(5.7)$$
where we have exhibited the resulting contributions at order $\mu, \mu^2$ and
$\mu^3$, in order. One sees that the series actually converges slowly for $\mu
\simeq 0.14$. Notice, however, that these results should be considered
indicative since higher loop corrections will modify the coefficients of the
$\mu^3 $ and $\mu^3 \ln \mu$ terms. The result of the full one loop calculation
will be given in section VI and compared to the recent experimental
determination [3]. In the charged channels the corrections are increasing with
with increasing $|k^2|$ due to form factor effects (see section 6). This is
different from the photoproduction case, where the loop corrections to
$E_{0+}^{\gamma p \to \pi^+ n}(\mu) $ and $E_{0+}^{\gamma n \to \pi^- p}(\mu)$
are small and move the chiral prediction closer to the empirical values.
\medskip
\noindent{\bf V.3. INCLUSION OF SOME ISOSPIN BREAKING EFFECTS}
\medskip
As already stated, the LETs presented so far refer strictly to the isospin
limit and do not account for isospin breaking either through quark mass
differences $(\sim m_d - m_u$) or (higher order) electromagnetism ($\sim e^2$).
At present, we are not able to implement in a consistent way all possible
effects of isospin breaking. However, to get an idea about the size of such
effects, we will present the results of a simplified approach. It is well-known
that the mass of neutral and charged pion differ slightly,
$${M_{\pi^+} - M_{\pi0}  \over  M_{\pi^+}}  = 0.033 \eqno(5.8)$$
and this difference is almost entirely of electromagnetic origin [29]. One
therefore expects some effect to come from this mass difference of pions in the
loop and the asymptotic state. Indeed, taking this mass difference into
account in neutral pion photoproduction leads to a substantial correction and
$E_{0+}^{\gamma p \to \pi^0 p} = - 1.97 \cdot 10^{-3}/M_{\pi^+}$ in good
agreement with the commonly accepted empirical value $E_{0+}^{\gamma p \to
\pi^0 p } = -(2.0 \pm 0.1)\cdot 10^{-3} / M_{\pi^+}$
[7].\footnote{$^*$}{In ref.[7], isospin breaking was only considered in
the class I diagrams which give the dominant loop--effect at $k^2 =0$.}
A similar mechanism is also
operative in the function $\Xi_{1,2}(\rho)$ which enter neutral pion
electroproduction. Expanding in $\mu_0 = M_{\pi^0} / m $ and using $\rho_0 = -
\nu /\mu_0^2$ we find
$$\eqalign{
\tilde \Xi_1(\rho_0) & = {4\over M_{\pi^0} } \biggl\{ - \sqrt{M^2_{\pi^+} -
M^2_{\pi^0} } + \int_0^1 dx \sqrt{M^2_{\pi^+} - x^2 M^2_{\pi^0} + k^2 x(x-1)}
\biggr\}\,,\cr \tilde \Xi_2(\rho_0) & = 2 \int_0^1 dx { x(1-2x) M_{\pi^0} \over
\sqrt{ M^2_{\pi^+} - x^2 M^2_{\pi^0} + k^2 x(x-1) }}\,. \cr} \eqno(5.9)$$
In fig.[3], we show the functions $\tilde \Xi_{1,2}(\rho_0)$. Their $k^2$
dependence is similar to the isospin symmetric ones $\Xi_{1,2}(\rho)$, but the
value at the photon point is substantially reduced. While when neglecting the
$\pi^+-\pi^0$ mass difference, $\Xi_1(0) = \pi\simeq 3.14 $ and $\Xi_2(0) =
2- \pi\simeq -1.14$,
we have now $\tilde \Xi_1(0) = 2.28$ and $\tilde \Xi_2(0) =
-0.74$,
which amount to a sizable reduction of $27\%$ and $35\%$, respectively.
The main point to stress here is that in neutral pion electroproduction it is
mandatory to go beyond ${\cal O}(q^3)$ and to include at least in a simplified
manner isospin breaking effects (like the charged and neutral pion mass
difference). A systematic study of these effects is beyond the scope of this
paper.

Differentiating between the neutral and charged pion masses in the
loops amounts to a redefinition of some loop functions as detailed
in appendix D. In essence, one has to disentangle the two thresholds,
$s_{\rm thr}^0 = m + M_{\pi^0}$ and
$s_{\rm thr}^+ = m + M_{\pi^+}$ ( the neutron-proton mass difference
is not accounted for). In particular, some of the low--energy constants
which enter the proton charge radius and anomalous magnetic moment
have to be readjusted. We find $b_9^r(1 {\rm GeV})
+ {\tilde b}_9^r(1 {\rm GeV}) = 5.17 \cdot 10^{-3}$ as compared to
$5.30 \cdot 10^{-3}$ in the isosymmetric case (see eq.(3.12)).
Similarly, $\delta \kappa_p$ defined in eq.(4.4) changes from 1.288
to 1.291. That this prescription conserves gauge invariance can be
checked after constructing the amplitudes $B_i (s,u)$ and using
eqs.(2.3). We should also stress that in this case of pion
electroproduction one has to include the pion mass difference in the
diagrams of all
classes (the procedure used in ref.[7] for
photoproduction is, however, legitimate).
\bigskip
\noindent{\bf VI. RESULTS AND DISCUSSION}
\medskip
\noindent{\bf VI.1. ISOSYMMETRIC CASE}
\medskip
First, we consider the isosymmetric case, i.e. we do not differentiate
between the proton and the neutron masses and also not between the
neutral and charged pion masses. Throughout, we set $m_p = m_n = m
= 938.27$ MeV, $F_\pi = 93.1$ MeV, $g_{\pi N}^2 / 4\pi = 14.28$
and $e^2 / 4\pi
= 1/ 137.036$. When considering charged pion production we set
$M_\pi = M_{\pi^\pm} = 139.57$ MeV
and for neutral pion production
$M_\pi = M_{\pi^0} = 134.97$ MeV. This is a consistent procedure
within the one--loop approximation we are using.

In Fig.4 we compare the CHPT prediction with the recent NIKHEF data
[3]. In the same figure, we also show the results from the tree
diagrams (open circles) and from the pseudovector (PV) Born terms
with form factors (open squares)  calculated in Ref.[8]. Note that
all of the PV results presented here do not include vector meson--exchange
and final state interaction  terms introduced in [8]. We see that the
one--loop contribution drastically reduces the calculated cross sections
and brings the CHPT results within the experimental uncertainties.
The results from the PV terms are also not too different from the data.
To have a more rigorous test of CHPT more precise measurements at small
$k^2$ are needed.

Let us now turn to analyzing the dynamical content of the CHPT calculations.
We consider a kinematical situation with the final pion--nucleon
invariant mass $W=1080 /1074$ MeV (for charged/neutral pion production)
which is perhaps most realistic for future experiments. At this near
threshold energy ($W_{{\rm thr}} = 1077.84/1073.24$ MeV for
charged/neutral production) the cross sections are dominated by the
S--wave amplitudes $E_{0+}$ and $L_{0+}$ ($S_{0+}$). In Figs.5 and 6
we display the real parts of these two amplitudes for $p (\gamma^\star, \pi^+)
n$, $n (\gamma^\star, \pi^-) p$ and
$p (\gamma^\star, \pi^0) p$, in order. In all cases the tree and PV results
are rather similar whereas the one--loop prediction is significantly
different from these. This comes in part from the irreducible
class I,II,III diagrams
and to a large extent from the difference in the $k^2$ dependence of the
pion form factor and the isovector nucleon charge form factor which is
a natural ingredient in our approach.\footnote{$^*$}{The role of the
various form factors has also been stressed by Olsson et al. in ref.[31].}
In Ref.[30] we had already shown
that the S--wave cross section for neutral pion production off the
proton
$$ a_0 = |E_{0+}|^2 - \epsilon {k^2 \over  k_0^2} |L_{0+}|^2
\eqno(6.1)$$ where $\epsilon$ and $k_0 = (s -m^2 +k^2)/2 \sqrt s$
represent, respectively, a measure of the transverse linear polarization
and the energy of the virtual photon in the $\pi N$ rest frame, follows
nicely the data of Welch {\it et al.} [3] whereas PV and tree are at
variance with the data with increasing $|k^2|$. In this channel we find
at the photon point ($k^2 = 0$) $|L_{0+}|^2 = 0.2 \, \mu b$ in fair
agreement with the result of [3],
$|L_{0+}|^2 = 0.13 \pm 0.05 \, \mu b$. Clearly, a precise measurement
of the $k^2$ dependence is the most effective way to test the CHPT
predictions. We should stress that one should consider $|k^2| \le
0.05$ GeV$^2$ since otherwise the loop corrections are too large.
To further illustrate this point, we show in Fig.7       the transverse
and longitudinal cross sections $\sigma_T$ and $\sigma_L$, respectively,
for $p (\gamma^\star, \pi^+)
n$, $n (\gamma^\star, \pi^-) p$ and
$p (\gamma^\star, \pi^0) p$, in order. In all cases the one--loop predictions
are rather different from the PV and tree results. We find an enhancement
of $\sigma_L$ and a depletion of $\sigma_T$. Note, however, that the
sums $\sigma = \sigma_T + \sigma_L$ are very similar for tree, PV and
one--loop.
To test the dynamical content
of the chiral symmetry prediction,  it is therefore mandatory to
perform a transverse/longitudinal separation.
To further exhibit these differences,
we plot in Figs.8 to 11 the differential cross sections
$d \sigma_T / d\Omega$,
$d \sigma_L / d\Omega$,
$d \sigma_I / d\Omega$ and
$d \sigma_P / d\Omega$
for three values of $k^2 = -0.001, -0.04$
and $-0.08$ GeV$^2$.
The most striking difference appears
in
$d \sigma_T / d\Omega$ for neutral pion production. While the CHPT
prediction becomes forward peaked for $|k^2| > 0.04$ GeV$^2$, the
PV and tree results remain backward peaked until
$|k^2| \simeq 0.07$ GeV$^2$. For
$|k^2| \le 0.05$ GeV$^2$ the shapes of
$d \sigma_I / d\Omega$ and
$d \sigma_P / d\Omega$ are not too different between the tree, loop
and PV predictions in case of charged pion electroproduction.
For neutral electropionproduction, the angular distribution
$d\sigma_P / d \Omega$ shows a significant difference between the
tree, PV and one loop results even at low photon four--momenta.
We will come back to this in section 6.2. It is worth to stress
that as $|k^2|$ increases, the contributions from the diagrams of
classes II and III become more and more important. To illustrate
this, consider the S--wave cross section (6.1) at $k^2 = -0.06$
GeV$^2$. The reduction of the one loop result compared to the tree
(or PV) prediction [30]
is almost entirely due to the class II and III
diagrams (with equal share). The class I diagrams contribute insignificantly
to $a_0$ at this $k^2$. This is completely different from the
photoproduction case discussed in refs.[1,7].

It is important to note that these results at $W = 1074/1080$ MeV
shown in Figs.5 to 11 contain contributions from P--wave multipoles.
In the threshold region these are small but they can nevertheless
contribute significantly to the various cross sections through the
interference with the S--wave amplitudes.  In table 1a,b,c we give the
real and imaginary parts for
$E_{0+}$,
$S_{0+}$,
$E_{1+}$,
$S_{1+}$,
$S_{1-}$,
$M_{1+}$ and
$M_{1-}$ for $k^2 = -0.001$, $-0.04$ and $-0.06$ GeV$^2$
for $p (\gamma^\star, \pi^+)
n$, $n (\gamma^\star, \pi^-) p$ and
$p (\gamma^\star, \pi^0) p$, in order. The knowledge of these multipoles
allows one to construct any observable one wishes to measure, in particular
the many which appear in the case of polarized electrons. Some
pertinent formulae are summarized in appendix C.

To explore the sensitivity of the charged pion production amplitudes to
the axial form factor, we have performed a set of calculations at
$k^2 = -0.04$ GeV$^2$ varying the dipole mass between 0.96 and 1.16 GeV,
corresponding to squared axial radii between 0.35 and 0.51 fm$^2$
(remember that our central value is $M_A = 1.06$ GeV). From the LETs
discussed before, eq.(5.1), one expects that a smaller axial radius
leads to an increase in transverse strength and a decrease of the
longitudinal strength.  This expectation is borne out by the actual
calculations. In fig.12     we show the differential cross sections
$d \sigma_T / d\Omega$,
$d \sigma_L / d\Omega$,
$d \sigma_I / d\Omega$ and
$d \sigma_P / d\Omega$ for this range of axial radii for $p (\gamma^\star,
\pi^+)n$
(the results for $n (\gamma^\star ,\pi^-)p$ are similar and not
shown). Particularly
sensitive to $M_A$ is the ratio $\sigma_T / \sigma_L$ which decreases/increases
by approximately 20 per cent for $M_A$ lowered/enhanced by 0.1 GeV. In
contrast, the sum $\sigma_L + \sigma_T$ is almost independent of the
axial radius which again points towards the importance of a
transverse/longitudinal separation. A summary of these results is given
in table 2.

Furthermore, we have investigated the influence of the counterterm
$d_1$ which already appears in the photoproduction reaction
$p(\gamma , \pi^o )p$. Reducing its strength
by a factor 2, the transverse cross section is diminished since the
electric dipole amplitude is smaller (at the photon point). For finite
$k^2$, however, this trend reverses and eventually leads to enhancement
of $\sigma_T$ (by 50 per cent at $k^2 = -0.04$ GeV$^2$). The longitudinal
cross section is essentially unaffected by this counter term. The charged
electroproduction amplitudes are insensitive to this finite contact term.
We should stress again that its central value could be considered as
coming from
a best fit  of
the precise total photoproduction cross section data [15]. We will
come back to this when discussing the inclusion of some isospin--breaking
effects. We did not vary the two counter terms appearing in the
charged photoproduction channels since their influence on the pertinent
observables is fairly small.

\medskip
\noindent{\bf VI.2. INCLUDING ISOSPIN--BREAKING}
\medskip
As discussed in section 5.3, for the case of neutral pion elelectroproduction
it is mandatory to consider the dominant isospin breaking effect which
stems from the pion mass difference. In what follows, we consider
$\gamma^\star p \to \pi^0 p$ using $M_{\pi^+} = 139.57$ MeV and
$M_{\pi^0} = 134.97$ MeV for the pertinent loop functions (see appendix
D). In all kinematical factors the neutral pion mass enters. Before
presenting the results, it is worth to stress that the gauge invariance
conditions (2.3) are fulfilled within machine accuracy.

In fig.13, the S--wave cross section $a_0$ defined in eq.(6.1) is shown
in comparison to the isosymmetric CHPT  [30] and the PV result (for
$\epsilon = 0.58$).
Although
the pion mass difference has a pronounced
effect on $E_{0+}(\mu , 0)$ and
$L_{0+}(\mu , 0)$ (e.g. $|L_{0+}|^2 = 0.17 \, \mu b$), at finite $k^2$
the two CHPT curves
do not differ significantly. The conclusion of ref.[30]
that loop effects are needed to explain the trend of the NIKHEF data
is not invalidated by the inclusion of isospin-breaking. This is due to
the dominating effect of the class II and class III diagrams at finite
$k^2$.

To get a better handle on the possible isospin-breaking effects, we show
in fig.14 the angular distributions $d\sigma_{L,T,P,I} / d\Omega$ at
$k^2 = -0.04$ GeV$^2$. Only in the case of $d\sigma_P / d\Omega$ one
finds a significant change from the isosymmetric case. The height in
the peak differs by almost a factor of three. In all other cases, the
differences are marginal. This can also be seen from the comparison of
$\sigma_T$ and $\sigma_L$ shown in fig.15 for $-0.1 \le k^2 \le 0$
GeV$^2$. The largest sensitivity of these results stems from the
low--energy constant $d_1$. Reducing its strength by a factor of two,
one finds an enhancement of $d\sigma_T / d\Omega$ in forward direction
(at $\theta = 0$, the enhancement factor is about 1.7) and of
$d\sigma_P / d\Omega$ (the peak height is increased by about 1.4).
The other two angular distributions remain almost unaltered. In the
S--wave cross section $a_0$ this amounts to an increase of about 21
percent. We conclude that the largest uncertainty of our predictions
stems from the knowledge of the low--energy constant $d_1$.

\baselineskip 12pt plus 1pt minus 1pt
\vskip 4truecm
\centerline{\bf ACKNOWLEDGEMENTS}
\medskip
We are grateful to Nimai Mukhopadhyay for some useful comments.
This work is partially supported by the U.S. Department of Energy,
Nuclear Physics Division, under contract W-31-109-ENG-38.
\bigskip \vfill \eject
\noindent{\bf APPENDIX A: INVARIANT AMPLITUDES OF PION ELECTROPRODUCTION}
\medskip
Here, we will complete the list of invariant amplitudes for pion
electroproduction by giving those for the reaction $\gamma^* n \to \pi^0 n$ as
well as the (-)   amplitudes.
\medskip
\noindent{\bf A.1. THE CHANNEL $\gamma^* n \to \pi^0 n$}
\medskip
\noindent{\underbar{Electric and magnetic Born terms:}}
$$\eqalign{
& A_1(s,u) = - B_2(s,u) = -eg_{\pi N} \biggl( {1\over s - m^2} + {1\over u -
m^2} \biggr)\, F_1^n(k^2)\,, \cr
& B_1(s,u)+2B_4(s,u) = A_{3,4,6}(s,u) = 0 \,.\cr } \eqno(A.1)$$
with the neutron Dirac form factor $F_1^n(k^2)$ given in eq.(3.12).
$$\eqalign{
A_1(s,u) = & -{e g_{\pi N} \over 2m^2} \, \delta \kappa_n , \qquad
A_{2,5,6}(s,u) = 0, \cr
A_3(s,u) = & {eg_{\pi N}\over 2 m} \delta \kappa_n
\biggl( {1\over s - m^2} - {1\over u - m^2}
\biggr), \cr A_4(s,u) = &  {eg_{\pi N} \over 2 m} \delta \kappa_n
\biggl( {1\over s - m^2} +
 {1\over u - m^2} \biggr)\,. \cr } \eqno(A.2)$$
where
$$\delta \kappa_n = \kappa_n + 16 g_{\pi N}^2 m^2 \bigl\{ \Gamma_4(m^2,0)+
\gamma_4(m^2,0) \bigr\}  = 1.821 \eqno(A.3)$$
is adjusted to reproduce the empirical value of the neutron anomalous magnetic
moment, $\kappa_n = -1.913$.
\medskip
For class I and II the following representation is again most economic
$$\eqalign{
A_1(s,u) & = a_1(s,k^2) + a_1(u,k^2) - B_2(s,u),\cr
B_2(s,u) & = {k^2 \over s - m^2} \bigl[ b_{14}(m^2,k^2) - b_{14}(s,k^2) \bigr]+
{k^2 \over u - m^2} \bigl[ b_{14}(m^2,k^2) - b_{14}(u,k^2) \bigr],
\cr B_1(s,u) + 2B_4(s,u) & = b_{14}(s,k^2) - b_{14}(u,k^2),\qquad
A_3(s,u)  = a_3(s,k^2) - a_3(u,k^2), \cr
A_4(s,u) & = a_3(s,k^2) + a_3(u,k^2),\qquad A_6(s,u) = a_6(s,k^2) -
a_6(u,k^2)\,.\cr} \eqno(A.4)$$
\medskip
\vfill
\eject
\noindent{\underbar{Class I:}}
\medskip
$$\eqalign{
a_1(s,k^2) & = {eg_{\pi N }\over F_\pi^2}  (   s-m^2) [ 2\gamma_6(s,k^2) -
2\gamma_4(s,k^2) + \Gamma_1(s,k^2) - \Gamma_2(s,k^2) - 2 \Gamma_4(s,k^2) \cr &
\qquad \qquad +2 \Gamma_6(s,k^2)     ], \cr
b_{14}(s,k^2) & =  {eg_{\pi N} \over F_\pi^2} \bigl\{   (s+3m^2) [ 2
\gamma_6(s,k^2) - \gamma_2(s,k^2)+ 2\Gamma_6(s,k^2) ] \cr & \qquad \qquad
+(s-m^2) [
\gamma_1(s,k^2) - 2 \gamma_5(s,k^2) +\Gamma_1(s,k^2) -\Gamma_2(s,k^2)
 -2\Gamma_5(s,k^2)] \bigr\}, \cr
a_3(s,k^2) & = {4eg_{\pi N} \over F_\pi^2} m [ \gamma_4(s,k^2)+
\Gamma_4(s,k^2)], \cr
a_6(s,k^2)  & = {2eg_{\pi N} \over F_\pi^2} m [ 2 \gamma_6(s,k^2) -
\gamma_2(s,k^2) +2\Gamma_6(s,k^2)]\,.\cr } \eqno(A.5)$$
\noindent{\underbar{Class II:}}
$$\eqalign{
a_1(s,k^2)  = {eg_{\pi N }^3\over 2m^2} & \bigl\{(s-m^2) [ 2\gamma_4(s,k^2) -
2\gamma_6(s,k^2)- \Gamma_1(s,k^2) +\Gamma_2(s,k^2) \cr & +2 \Gamma_4(s,k^2) - 2
 \Gamma_6(s,k^2)  ]  + 8m^2 [\gamma_4(s,k^2) + \Gamma_4(s,k^2) ]\bigr\},\cr
b_{14}(s,k^2)  = {eg_{\pi N}^3 \over 2m^2} & \bigl\{ (s+3m^2) [2\gamma_5(s,k^2)
-  \gamma_1(s,k^2)- \Gamma_1(s,k^2) + \Gamma_2(s,k^2) + 2 \Gamma_5(s,k^2) ]
\cr & +{s^2 + 10 sm^2 + 5m^4 \over s-m^2}  [  \gamma_2(s,k^2) -
2\gamma_6(s,k^2)- 2 \Gamma_6(s,k^2)  ] \bigr\} ,\cr
a_3(s,k^2)  = {eg_{\pi N}^3 \over m} & \bigl\{ 2\gamma_6(s,k^2) - 4
\gamma_4(s,k^2)- 4\Gamma_4(s,k^2) + \Gamma_1(s,k^2) -  \Gamma_2(s,k^2) \cr & +
2 \Gamma_6(s,k^2) - {8m^2 \over s - m^2} [ \Gamma_4(s,k^2) +
\gamma_4(s,k^2) ]\bigr\}, \cr
a_6(s,k^2)   = {eg_{\pi N}^3 \over m} & \bigl\{  2 \gamma_5(s,k^2)
-\gamma_1(s,k^2) - \Gamma_1(s,k^2) + \Gamma_2(s,k^2)
+2\Gamma_5(s,k^2) \cr & + {2(s+m^2) \over s - m^2} [ \gamma_2(s,k^2) -
2\gamma_6(s,k^2) -2 \Gamma_6(s,k^2)    ] \bigr\}\,.\cr } \eqno(A.6)$$
\vfill
\eject
\noindent{\underbar{Class III:}}

We give the $B_i(s,u)$ with $i\ne 3,5$ which allow uniquely to construct the
$A_i(s,u)$.
$$\eqalign{B_1(s,u)  = & eg_{\pi N}^3 \biggl\{
2(s-m^2) G_1(s,t,k^2) - 4 G_4(s,t,k^2)  - 2k^2 G_5(s,t,k^2) \cr &
+2(u-s) G_6(s,t,k^2) + 2(t+k^2 - M_\pi^2 ) G_7(s,t,k^2) + 2(t-4m^2)
G_8(s,t,k^2) \cr  & - 2t G_9(s,t,k^2) + 2 \Omega_4(s,u,k^2)  +
{s-m^2 \over 2m^2} \bigl[2\gamma_4(s,k^2)- 2\gamma_6(s,k^2) -\Gamma_1(s,k^2)
\cr & + \Gamma_2(s,k^2)  + 2 \Gamma_4(s,k^2) - 2 \Gamma_6(s,k^2) \bigr]
+ {k^2 \over 2m^2} \bigl[- \gamma_1(s,k^2)  +
 \gamma_2(s,k^2) + 2\gamma_5(s,k^2) \cr & - 2\gamma_6(s,k^2)  -\Gamma_1(s,k^2)
+  \Gamma_2(s,k^2) + 2 \Gamma_5(s,k^2) - 2 \Gamma_6(s,k^2)
\bigr]  \biggr\} + (s \leftrightarrow u)\,, \cr } $$

$$\eqalign{B_2(s,u)  = & eg_{\pi N}^3 \biggl\{  2 \tilde \Gamma_1(t) +
(s-m^2) \bigl[ 2G_2(s,t,k^2)-  2G_3(s,t,k^2) + \Omega_2(s,u,k^2)
\cr &
- 2 \Omega_6(s,u,k^2) \bigr]   + 4G_4(s,t,k^2) + 2k^2 G_5(s,t,k^2) +
2(2m^2-s-u) G_6(s,t,k^2)
\cr &
+ 2( M_\pi^2 - t - k^2)
G_7(s,t,k^2)  + 2 t G_9(s,t,k^2) +2(u-s-4m^2+M_\pi^2 - k^2) G_8(s,t,k^2)
\cr &
+ 2(s-m^2 - M_\pi^2) \bigl[2G_{10}(s,t,k^2) + \Omega_{10}(s,u,k^2) \bigr]
 + 2(M_\pi^2 - 3 m^2 - s) \Omega_8(s,u,k^2) \cr & - 2\Omega_4(s,u,k^2) +
 2\gamma_2(s,k^2) + {k^2 \over 2m^2} \bigl[ \gamma_1(s,k^2) - \gamma_2(s,k^2)
 -2\gamma_5(s,k^2) \cr &
+ 2\gamma_6(s,k^2) +\Gamma_1(s,k^2)  - \Gamma_2(s,k^2) - 2 \Gamma_5(s,k^2)
+ 2 \Gamma_6(s,k^2) \bigr] \biggr\} + (s \leftrightarrow u)\,, \cr } $$

$$\eqalign{B_1(s,u)  + & 2B_4(s,u) = eg_{\pi N}^3 \biggl\{
(s-m^2) \bigl[ 2G_1(s,t,k^2)+ 2G_2(s,t,k^2) - 2G_3(s,t,k^2) \cr & - 4
G_5(s,t,k^2) + \Omega_1(s,u,k^2) - \Omega_3(s,u,k^2) -2\Omega_5(s,u,k^2) \bigr]
+ 2( M_\pi^2 - 3 m^2 -s)\cr & \cdot \bigl[ 2G_6(s,t,k^2)  - 2G_{10}(s,t,k^2)
+\Omega_6(s,u,k^2) - \Omega_{10}(s,u,k^2) \bigr]   + 2(2s-2m^2-M_\pi^2) \cr &
\cdot\bigl[ 2G_7(s,t,k^2)  + \Omega_7(s,u,k^2) \bigr] + 2(m^2 + M_\pi^2 -s)
\bigl[ 2G_9(s,t,k^2) + \Omega_9(s,u,k^2) \bigr] \cr & + \gamma_0(s,k^2)-{s+3m^2
\over 2 m^2} \gamma_1(s,k^2)    + {s-m^2 \over 2 m^2} \bigl[ - \Gamma_1(s,k^2)
 + \Gamma_2(s,k^2) + 2\Gamma_5(s,k^2)  \cr & -  2
\Gamma_6(s,k^2) +  \gamma_2(s,k^2) +  2 \gamma_5(s,k^2) - 2
\gamma_6(s,k^2) \bigr] \biggr\} - (s \leftrightarrow u)\,, \cr } $$

$$\eqalign{ B_6(s,u) =  {2e\over m}  g_{\pi N}^3 \bigl\{ & \Gamma_1(s,k^2) -
\Gamma_2(s,k^2) + 2 \Gamma_6(s,k^2) + 2\gamma_6(s,k^2) \cr & - 8m^2
G_8(s,t,k^2) - 4m^2 \Omega_8(s,u,k^2) \bigr\} + (s \leftrightarrow u)\,,\cr} $$

$$\eqalign{ B_7(s,u) = &{e\over m}  g_{\pi N}^3 \bigl\{   \Gamma_1(s,k^2) -
\Gamma_2(s,k^2) - 2 \Gamma_5(s,k^2) + \gamma_1(s,k^2)  - 2\gamma_5(s,k^2) \cr &
+ 4m^2 [-2G_6(s,t,k^2) + 2 G_{10}(s,t,k^2) - \Omega_6(s,u,k^2) +
  \Omega_{10}(s,u,k^2) ]\bigr\} - (s \leftrightarrow u)\,, \cr } $$

$$\eqalign{ B_8(s,u) = & {e \over m}  g_{\pi N}^3 \bigl\{   \Gamma_1(s,k^2) -
\Gamma_2(s,k^2) + 2 \Gamma_6(s,k^2) + 2\gamma_6(s,k^2) \cr & - 8m^2
G_{10}(s,t,k^2) - 4m^2 \Omega_{10}(s,u,k^2) \bigr\} - (s \leftrightarrow u)\,.
\cr } \eqno(A.7)$$
\medskip
\noindent{\bf A.2. THE (-)   AMPLITUDES}
\medskip
\noindent{\underbar{Electric and magnetic Born terms:}}
$$\eqalign{
& A_1^{(-)}(s,u) = - B_2^{(-)}(s,u) = {e\over 2}g_{\pi N} \biggl( {1\over s -
m^2} -  {1\over u - m^2} \biggr)\, F_1^v(k^2)\,, \quad A_{3,4}^{(-)}(s,u) = 0
\,,\cr & B_1^{(-)}(s,u) +2 B_4^{(-)}(s,u) = {eg_{\pi N} \over t-M_\pi^2 } \,
F_\pi^v(k^2)
+ {eg_{\pi N} \over k^2} \, [ F_1^v(k^2) - F^v_\pi(k^2) ] \,, \cr &
A_6^{(-)}(s,u) = {eg_{\pi N} \over 2m k^2} [ F_1^v(k^2) - 1] \cr }\eqno(A.8)$$
Here, $F_1^v(k^2) = F_1^p(k^2) - F_1^n(k^2) $ denotes the nucleon isovector
Dirac form factor and
$$F^v_\pi(k^2) = 1 + {k^2 \over 6}  <r^2>_\pi - {2\over F_\pi^2} \,  \dbar{
J}^{\pi\pi}_{22}(k^2) \eqno(A.9)$$
is the one loop representation of the
pion charge form factor. It is important to note  that the Born amplitude is
more than just the tree amplitude (with form factors equal to unity) multiplied
by the appropriate form factors  generated by the loops. It is known that such
a simple multiplication prescription violates  gauge invariance, unless
the (unphysical) constraint $F_1^v(k^2) = F^v_\pi(k^2)$ is imposed (or
one adds some additional seagull terms [2,16]).
Within the
framework of CHPT the form factors generated by loops and counterterms are
automatically included in a gauge invariant fashion.
$$\eqalign{
A^{(-)}_{1,2,5,6}(s,u) & = 0, \cr
A_3^{(-)}(s,u) = &- {eg_{\pi N}\over 4 m} \delta \kappa_v
\biggl( {1\over s - m^2} + {1\over u - m^2}
\biggr), \cr A_4^{(-)}(s,u) = & - {eg_{\pi N} \over 4 m} \delta \kappa_v
\biggl( {1\over s - m^2} -
 {1\over u - m^2} \biggr)\,. \cr } \eqno(A.10)$$
where $\delta \kappa_v =  \delta \kappa_p - \delta \kappa_n$
is adjusted to reproduce the empirical value of the nucleon isovector anomalous
magnetic moment, $\kappa_v = 3.71$.
\medskip
For class I and II the following representation of the (-)   amplitudes is most
economic,
$$\eqalign{
A_1^{(-)}(s,u) & = a_1(s,k^2) - a_1(u,k^2) - B_2^{(-)}(s,u),\cr
B^{(-)}_2(s,u) & = -{k^2 \over s - m^2}\, b_{14}(s,k^2)+
{k^2 \over u - m^2} \, b_{14}(u,k^2) \,,
\cr B_1^{(-)}(s,u) + 2B_4^{(-)}(s,u) & = b_{14}(s,k^2) + b_{14}(u,k^2),\qquad
A^{(-)}_3(s,u)  = a_3(s,k^2) + a_3(u,k^2), \cr
A^{(-)}_4(s,u) & = a_3(s,k^2) - a_3(u,k^2),\qquad A^{(-)}_6(s,u) = a_6(s,k^2) +
a_6(u,k^2)\,.\cr} \eqno(A.11)$$
\medskip
\noindent{\underbar{Class I:}}
$$\eqalign{a_1(s,k^2) & = {eg_{\pi N }\over F_\pi^2} (s-m^2) [ \gamma_6(s,k^2)
- \gamma_4(s,k^2) ], \cr
b_{14}(s,k^2) & = {eg_{\pi N} \over 2F_\pi^2} \bigl\{ (s+3m^2) [ 2
\gamma_6(s,k^2) - \gamma_2(s,k^2) ] +(s-m^2) [ \gamma_1(s,k^2) - 2
\gamma_5(s,k^2) ] \bigr\}, \cr
a_3(s,k^2) & = {2eg_{\pi N} \over F_\pi^2} m \gamma_4(s,k^2), \cr
a_6(s,k^2)  & = {eg_{\pi N} \over F_\pi^2} m [ 2 \gamma_6(s,k^2) -
\gamma_2(s,k^2) ]\,.\cr } \eqno(A.12)$$
\medskip
\noindent{\underbar{Class II:}}
$$\eqalign{
a_1(s,k^2)   = {eg_{\pi N }^3\over 8m^2}\bigl\{ & (s-m^2) [ 8\gamma_4(s,k^2) -
8\gamma_6(s,k^2)- \Gamma_1(s,k^2) + \Gamma_2(s,k^2) \cr & +2 \Gamma_4(s,k^2)
 -2 \Gamma_6(s,k^2)  ]  + 8m^2 [4\gamma_4(s,k^2) + \Gamma_4(s,k^2) ]\bigr\},\cr
b_{14}(s,k^2)  = {eg_{\pi N}^3 \over 8m^2} \bigl\{ & (s+3m^2) [8\gamma_5(s,k^2)
- 4 \gamma_1(s,k^2)- \Gamma_1(s,k^2) + \Gamma_2(s,k^2) + 2 \Gamma_5(s,k^2) ]
\cr & +{s^2 + 10 sm^2 + 5m^4 \over s-m^2}  [  4 \gamma_2(s,k^2) -
8\gamma_6(s,k^2)- 2 \Gamma_6(s,k^2)  ] + 4m^2[-8\tilde\gamma_5(k^2) \cr & + 4
\tilde\gamma_1(k^2) +\tilde\Gamma_1(k^2)-2\tilde\Gamma_5(k^2)
+8m^2(-2\gamma_2'(k^2) +4\gamma_6'(k^2) + \Gamma_6'(k^2))] \bigr\},
\cr
a_3(s,k^2)  = {eg_{\pi N}^3 \over 4m}  \bigl\{ & [ 8\gamma_6(s,k^2) - 16
\gamma_4(s,k^2)-4\Gamma_4(s,k^2) + \Gamma_1(s,k^2) \cr & -  \Gamma_2(s,k^2) +
2 \Gamma_6(s,k^2)   - {8m^2 \over s - m^2} [ \Gamma_4(s,k^2) +4
\gamma_4(s,k^2) ]\bigr\}, \cr
a_6(s,k^2)   = {eg_{\pi N}^3 \over 4m}  \bigl\{ & 8 \gamma_5(s,k^2)
-4\gamma_1(s,k^2) - \Gamma_1(s,k^2) + \Gamma_2(s,k^2)
+2\Gamma_5(s,k^2) \cr & + {4(s+m^2) \over s - m^2} [ 2\gamma_2(s,k^2) -
4\gamma_6(s,k^2) - \Gamma_6(s,k^2)]  -8\tilde \gamma_5(k^2) + 4 \tilde
\gamma_1(k^2) \cr & + \tilde\Gamma_1(k^2) -2 \tilde\Gamma_5(k^2) +8m^2[-2
\gamma_2'(k^2)+4\gamma_6'(k^2) + \Gamma_6'(k^2) ] \bigr\}\,.\cr } \eqno(A.13)$$
Here, the prime on the loop functions denotes the partial derivative with
respect to $s$ evaluated at $s=m^2$.  Note, that we have $b_{14}(m^2,k^2) =
 a_6(m^2,k^2) = 0$ such
that $B_2^{(-)}(s,u)$ has no more pole at $s=m^2$ or $u=m^2$. The residue at
these poles is proportional to the isovector Dirac form factor, already
contained in the electric Born term. Similarly, $A_6^{(-)}(m^2,m^2) = 0$.
\medskip
\noindent{\underbar{Class III:}}
\medskip
$$\eqalign{B_1^{(-)}(s,u)  = & {e\over 2}g_{\pi N}^3 \biggl\{
\Gamma_0^{\pi N}(s) - {M_\pi^2 \over s - m^2} \underline \Gamma^{\pi N}_0(s) -
\Gamma_2^{\pi N}(s) +
(m^2-s) G_1(s,t,k^2) + 2 G_4(s,t,k^2) \cr & + k^2 G_5(s,t,k^2)
+(s-u) G_6(s,t,k^2) + ( M_\pi^2 - k^2 -t) G_7(s,t,k^2)  + t G_9(s,t,k^2) \cr &
+ (4m^2-t)  G_8(s,t,k^2) + {s-m^2 +k^2 \over 4m^2 } \bigl[\Gamma_1(s,k^2) -
\Gamma_2(s,k^2)+ 2 \Gamma_6(s,k^2) \bigr] \cr & + {m^2 - s \over 2m^2}
\Gamma_4(s,k^2)  - {k^2 \over 2 m^2} \Gamma_5(s,k^2)
 \biggr\} - (s \leftrightarrow u)\,, \cr } $$

$$\eqalign{B_2^{(-)}(s,u)  = & {e\over 2}g_{\pi N}^3 \biggl\{  -\Gamma_0^{\pi
N}(s) + {M_\pi^2 \over s - m^2} \underline \Gamma^{\pi N}_0(s) +  \Gamma_2^{\pi
N}(s)  +(m^2-s) \bigl[ G_2(s,t,k^2)  \cr & - G_3(s,t,k^2)] -2G_4(s,t,k^2) - k^2
G_5(s,t,k^2) +(s+u-2m^2)  G_6(s,t,k^2)  \cr & + ( t+k^2-M_\pi^2)
G_7(s,t,k^2)  + (s-u+4m^2-M_\pi^2 + k^2) G_8(s,t,k^2)\cr & - t G_9(s,t,k^2) +
2(m^2 + M_\pi^2-s) G_{10}(s,t,k^2)
+ {k^2 \over 4m^2} \bigl[ -\Gamma_1(s,k^2) \cr & + \Gamma_2(s,k^2) + 2
\Gamma_5(s,k^2)  - 2 \Gamma_6(s,k^2) \bigr] \biggr\} - (s \leftrightarrow
u)\,, \cr } $$

$$\eqalign{B_1^{(-)}(s,u) + & 2B_4^{(-)}(s,u) = {e\over 2}g_{\pi N}^3 \biggl\{
(s-m^2) \bigl[ -G_1(s,t,k^2)- G_2(s,t,k^2) + G_3(s,t,k^2) \cr & + 2
G_5(s,t,k^2) \bigr] +  2(s -  M_\pi^2 + 3 m^2)\bigl[ G_6(s,t,k^2)  -
G_{10}(s,t,k^2) \bigr] \cr &  + 2(2m^2+M_\pi^2-2s) G_7(s,t,k^2) + 2(s-m^2 -
M_\pi^2) G_9(s,t,k^2) \cr & +
 {s-m^2 \over 4 m^2} \bigl[  \Gamma_1(s,k^2) - \Gamma_2(s,k^2) -
 2\Gamma_5(s,k^2) +  2 \Gamma_6(s,k^2) \bigr] \cr &- \tilde \Gamma_1(t) +
 {1\over t- M_\pi^2}\bigl[ (t-4m^2) \tilde \Gamma_1(t)+(4m^2 -M_\pi^2) \tilde
 \Gamma_1(M_\pi^2) \bigr] \biggr\}  + (s \leftrightarrow u)\,, \cr } $$

$$B_6^{(-)}(s,u) =  {e\over 2m}  g_{\pi N}^3 \bigl\{
-\Gamma_1(s,k^2) + \Gamma_2(s,k^2) - 2 \Gamma_6(s,k^2) +  8m^2
G_8(s,t,k^2) \bigr\} - (s \leftrightarrow u)\,, $$

$$\eqalign{ B_7^{(-)}(s,u) = {e\over 4 m}  g_{\pi N}^3 \bigl\{ &
-\Gamma_1(s,k^2) + \Gamma_2(s,k^2) + 2 \Gamma_5(s,k^2)
 + 8m^2 G_6(s,t,k^2) \cr & - 8m^2  G_{10}(s,t,k^2)\bigr\}
\qquad + (s
\leftrightarrow u)\,, \cr } $$

$$ B^{(-)}_8(s,u) =  {e\over 4m}  g_{\pi N}^3 \bigl\{ -\Gamma_1(s,k^2) +
\Gamma_2(s,k^2) - 2 \Gamma_6(s,k^2) + 8m^2 G_{10}(s,t,k^2)\bigr\} + (s
\leftrightarrow u)\,.  \eqno(A.14)$$
It is important to note that in the chiral limit $M_\pi = 0$ one has the
relation
$$B_7^{(-)}(s=u=m^2, t=k^2) = {e g_{\pi N} \over 2mt} \biggl[ 1 - {1\over g_A}
G_A(t) \biggr] \eqno(A.15)$$
with $G_A(t)$ the one loop expression for the nucleon axial form factor. This
relation contains the basic current algebra statement, that the nucleon axial
form factor enters the pion electroproduction amplitudes. Since in general
current algebra statements are exact only in the chiral limit we can not expect
eq.(A.15) to hold also for finite pion mass. In that case corrections beyond
current algebra show up  which are not constrained a priori.
\bigskip
\noindent{\bf APPENDIX B: LOOP FUNCTIONS AND THEIR IMAGINARY PARTS}
\medskip
Here, we will heavily borrow from appendix B of ref.[1] where all the loop
function occuring in photoproduction have been defined and given in terms of
their Feynman parameter representations. We will write down here only the
relevant extensions to $k^2 \ne 0$.
The one loop expressions of the form factors involve the loop function
$$\dbar { J}^{\pi\pi}_{22} (k^2 ) = -{1\over 32 \pi^2} \biggl\{ {k^2 \over
6} + \int_0^1 dx [ M_\pi^2 + k^2 x(x-1)] \, \ln \bigg[ 1 + {k^2 \over M_\pi^2 }
x(x-1) \biggr] \biggr\}    \eqno(B.1)$$
Furthermore, we have in our electroproduction amplitudes
$$\gamma_i(s,k^2)  = {1\over 16\pi^2} \int_0^1dx\int_0^1dy {(1-y)
p_i(x,y) \over h_\gamma(x,y;s,k^2)} \qquad (i\ne 3),\eqno(B.2)$$
with $h_\gamma(x,y;s,k^2) = M_\pi^2(1-y) + m^2 y^2 + (s-m^2)xy(y-1) +
k^2(1-y)^2x(x-1)$ and the $p_i(x,y)$ are the same as in ref.[1].
An other set of new functions is:
$$\Gamma_i(s,k^2) = {1\over 16\pi^2} \int_0^1dx\int_0^1dy {y
q_i(x,y) \over h_\Gamma(x,y;s,k^2) }\qquad (i\ne 3),\eqno(B.3)$$
with $h_\Gamma(x,y;s,k^2) = M_\pi^2(1-y) + m^2 y^2 + (s-m^2)xy(y-1) +
k^2y^2x(x-1)$ and the $q_i(x,y)$ are the same as in ref.[1]. For the nucleon
form factors their values  at $s=m^2$ occur which we denote by a tilde,
$$\eqalign{
\tilde \gamma_i(k^2) & = \gamma_i(m^2,k^2)\,, \qquad \tilde \Gamma_i(k^2) =
\Gamma_i(m^2,k^2)\,, \quad (i\ne 3), \cr
\bar{\tilde \gamma}_3(k^2) & = {1\over 32 \pi^2} \int_0^1 dx \int_0^1 dy (1-y)
\ln {h_\gamma(x,y;m^2,k^2) \over h_\gamma(x,y;m^2,0)} \,, \cr
\bar{\tilde \Gamma}_3(k^2) & = {1\over 32 \pi^2} \int_0^1 dx \int_0^1 dy y
\ln {h_\Gamma(x,y;m^2,k^2) \over h_\Gamma(x,y;m^2,0)} \,, \cr}\eqno(B.4)$$
and an overbar means subtraction at $k^2  =0$.
\medskip

The box graphs give rise to the $G$-functions,
$$\eqalign{
G_i(s,t,k^2) & = {1\over 16 \pi^2} \int_0^1 dx \int_0^1 dy \int_0^1 dz{(1-x)
y^2\,b_i(x,y,z) \over h_G^2(x,y,z;s,t,k^2)} \,, \quad (i\ne 4),\cr
G_4(s,t,k^2) & = {1\over 32 \pi^2} \int_0^1 dx \int_0^1 dy \int_0^1 dz{(x-1
)y^2\over h_G(x,y,z;s,t,k^2)} \,, \cr}\eqno(B.5)$$
with
$$\eqalign{  h_G(x,y,z;s,t,k^2)= & M_\pi^2[1-y+x(x-1)y^2z] +m^2 y^2 + (s-m^2)
xy (y-1) \cr & + t(1-x)^2 y^2 z(z-1) + k^2 x(x-1) y^2 (1-z)\cr}\eqno(B.6)$$
and the $\Omega$-functions,
$$\eqalign{
\Omega_i(s,u,k^2) & = {1\over 16 \pi^2} \int_0^1 dx \int_0^1 dy \int_0^1 dz
{(1-y)y\,r_i(x,y,z) \over h_\Omega^2(x,y,z;s,u,k^2)} \,, \quad (i\ne 4),\cr
\Omega_4(s,u,k^2) & = {1\over 32 \pi^2} \int_0^1 dx \int_0^1 dy \int_0^1 dz{(y
-1 )y \over h_\Omega(x,y,z;s,u,k^2)} \,, \cr}\eqno(B.7)$$
with
$$\eqalign{h_\Omega(x,y,z;s,u,k^2) = & M_\pi^2 [1-y+x(x-1)y^2] +m^2 y^2 +
(s-m^2) (1-x)y (y-1)z \cr & + (u-m^2) x y(1-y)(z-1) + k^2 (1-y)^2 z(z-1)\,.
\cr }  \eqno(B.8)$$
The polynomials $b_i(x,y,z)$ and $r_i(x,y,z)$ are the same as in ref.[1].
\medskip
Finally, we give the expressions for the imaginary parts of the new
$k^2$-dependent scalar loop functions (those with $i=0$). The imaginary part of
the vector and tensor functions are obtained in the standard way. One has
to write down appropriate linear relations among these functions. The solution
of a  subset of these linear equations with maximal rank gives rise to explicit
formulae  for the imaginary parts of the vector and tensor functions expressed
through those of the scalar ones.

The quadratic polynomial $\lambda(x,y,z) = x^2 + y^2 +z^2 - 2xy -2xz -2 yz$ is
called the K\"all\'en or triangle function. Let us abbreviate $\lambda_\pi =
\lambda(s,m^2,M_\pi^2)$ and $\lambda_k = \lambda(s,m^2,k^2)$. Using the
Cutkosky cutting rules we obtain the following imaginary parts:
$$
{\rm Im} \gamma_0(s,k^2)  = {1\over 16 \pi \sqrt \lambda_\pi} \ln
 {(s-m^2 +M_\pi^2) (s-m^2+k^2) - 2s k^2 + \sqrt{\lambda_\pi \lambda_k} \over
 (s-m^2 +M_\pi^2) (s-m^2+k^2) - 2s k^2 - \sqrt{\lambda_\pi \lambda_k} } \,,$$

$${\rm Im} \Gamma_0(s,k^2)  = {1\over 16 \pi \sqrt \lambda_\pi} \ln
 {(s+m^2 -M_\pi^2) (s-m^2+k^2) - 2s k^2 + \sqrt{\lambda_\pi \lambda_k} \over
(s+m^2 -M_\pi^2) (s-m^2+k^2) - 2s k^2 - \sqrt{\lambda_\pi \lambda_k} }  \,,$$

$$\eqalign{
{\rm Im} G_0(s,t,k^2) & =   { \sqrt \lambda_\pi \over 32 \pi \sqrt{B_G^2 - A_G
C_G}}  \ln { A_G + B_G +\sqrt{ B_G^2 - A_G C_G} \over A_G + B_G - \sqrt{ B^2_G
- A_G C_G} }  \,,\cr
A_G(s,t,k^2) & = m^2 s^2 -s( 2m^4 +k^2 M_\pi^2) + m^6 +k^2 M_\pi^2(k^2 +M_\pi^2
-3m^2) \,,\cr
B_G(s,t,k^2) & = -{t\over 2} \lambda_\pi + {1\over 2} M_\pi^2 (M_\pi^2 - k^2) (
M_\pi^2 +2 k^2 - 3m^2 -s) \,, \cr
C_G(s,t,k^2) & =  t \lambda_\pi + M_\pi^2 (M_\pi^2 - k^2 )^2 \,, \cr
} \eqno(B.9)$$

$$\eqalign{
{\rm Im} \Omega_0(s,u,k^2) & =  { \sqrt \lambda_\pi \over 32 \pi
\sqrt{B_\Omega^2 - A_\Omega C_\Omega}}  \ln { A_\Omega + B_\Omega +\sqrt{
B_\Omega^2 - A_\Omega C_\Omega} \over A_\Omega + B_\Omega - \sqrt{ B^2_\Omega
- A_\Omega  C_\Omega} }  \,,\cr
A_\Omega(s,u,k^2) & = M_\pi^2 (s-m^2)^2 +k^2 M_\pi^2( M_\pi^2 - 3m^2 -s) + k^4
m^2   \,,\cr
B_\Omega(s,u,k^2) & = -{u\over 2} \lambda_\pi + {s^2\over 2}(m^2 - M_\pi^2) +
s(-m^4 +2m^2 M_\pi^2 - {3\over 2} M_\pi^4 - m^2 k^2 + {3\over 2}  M_\pi^2 k^2 )
\cr &\quad  + {1\over 2} m^6
-{7\over 2} m^4 M_\pi^2 + 2 m^2 M_\pi^4 + k^2(m^4 +{5\over 2} m^2 M_\pi^2 -
{3\over 2} M_\pi^4 ) - k^4 m^2 \,, \cr
C_\Omega(s,u,k^2) & = u \lambda_\pi + 2s(M_\pi^2 -m^2) (M_\pi^2 - k^2 ) + (2m^2
-M_\pi^2 - k^2 )( 3m^2 M_\pi^2 - 2 M_\pi^4 - k^2 m^2)   \,. \cr} $$
\medskip
The formulae given above hold for $k^2 \le 0, \quad t \le 0, \quad u\le
(m-M_\pi)^2$ and $s \ge (m+M_\pi)^2$. The physical region is completely
contained within this domain of the Mandelstam plane.
\bigskip
\noindent{\bf APPENDIX C: CALCULATION OF CURRENT MATRIX ELEMENTS FROM
MULTIPOLE AMPLITUDES}
\medskip
To calculate polarization observables using the formulas of Ref.9, we
need to construct current matrix elements in the center-of-mass frame of
the final $\pi N$ system. The photon momentum
defines the z-axis
and hence we have $k^{\mu} = (k_0, 0, 0, k)$.
The amplitude of the $\gamma^{*} N \rightarrow \pi N$ process can then be
written as
$$
<m_{s^{\prime}} \mid \epsilon^{\mu}(\lambda) J_{\mu} \mid m_s > =
\sum_{i=1,6}<m_{s^{\prime}} \mid O_i \mid m_s > f_i(k^2,W,x) \eqno(C.1)
$$
with
$$\eqalign{
O_1 & = i \vec{\sigma}\cdot\vec{\epsilon}_{\lambda} \cr
O_2 & =\vec{\sigma}\cdot\hat{q}\vec{\sigma}\cdot(\hat{k}
\times \vec{\epsilon}_{\lambda}) \cr
O_3 & = i \vec{\sigma} \cdot \hat{k} \hat{q} \cdot \vec{\epsilon}_{\lambda} \cr
O_4 & = i \vec{\sigma} \cdot \hat{q} \hat{q} \cdot \vec{\epsilon}_{\lambda} \cr
O_5 & = i \vec{\sigma} \cdot \hat{k} \hat{k} \cdot \vec{\epsilon}_{\lambda} \cr
O_6 & = i \vec{\sigma} \cdot \hat{k} \hat{k} \cdot \vec{\epsilon}_{\lambda}
\cr}
\eqno(C.2)$$
Here we have defined $\vec{q}$ as the pion momentum in $\pi N$ center-of-mass
frame, $\hat{q} =\vec{q}/| \vec q|$
and $\hat{k}=\vec{k}/| \vec k|$. The spherical unit
vectors are defined as $\vec{\epsilon}_{\pm 1} = \mp 1/\sqrt{2} ( \hat{x}
\pm i \hat{y})$ and $\vec{\epsilon}_{0} = \hat{z}$. The photon polarization
vectors for $\lambda = \pm 1,0$ in Eq.(C.1) are defined as
$$\eqalign{
\epsilon^{\mu}(\pm 1) & = (0,\vec{\epsilon}_{\pm 1}) \cr
\epsilon^{\mu}(0) & = {1 \over \sqrt{-k^2}}
(| \vec k|, k_0 \vec{\epsilon}_{0} ) \cr}
\eqno(C.3) $$
where $k^2=k_0^2 - \vec{k}^2 < 0 $.
The coefficients $f_i(k^2,W,x)$ are only functions of $\pi N$ invariant mass
$W$ , four-momentum transfer square $k^2 $
and $ x =\hat{k}\cdot\hat{q}$.
In terms of the multipole amplitudes, we
have (all multipoles are functions of $k^2$ and $W$)
$$\eqalign{
f_1(k^2,W,x) & = \sum_{l}(E_{l+}P^{\prime}_{l+1}(x)
              + E_{l-}P^{\prime}_{l-1}(x)
          +M_{l+}P^{\prime}_{l+1}(x) +
         (l+1)M_{l-}P^{\prime}_{l-1}(x)) \cr
f_2(k^2,W,x) & = \sum_{l}((l+1)M_{l+}P^{\prime}_{l}(x)
     +l M_{l-}P^{\prime}_{l}(x)) \cr
f_3(k^2,W,x) & = \sum_{l}(E_{l+}P^{\prime\prime}_{l+1}(x)
+E_{l-}P^{\prime\prime}_{l-1}(x)
-M_{l+}P^{\prime\prime}_{l+1}(x)-M_{l-}P^{\prime\prime}_{l-1}(x)) \cr
f_4(k^2,W,x) & = \sum_{l}(-E_{l+}P^{\prime\prime}_{l}(x)
-E_{l-}P^{\prime\prime}_{l}(x) + M_{l+} P^{\prime\prime}_{l}(x)
-M_{l-}P^{\prime\prime}_{l}(x)) \cr
f_5(k^2,W,x) & = \sum_{l}(-(l+1)S_{l+}P^{\prime}_{l}(x)
+l S_{l-}P^{\prime}_{l}(x)) \cr
f_6(k^2,W,x) & = \sum_{l}((l+1)S_{l+}P^{\prime}_{l+1}(x)
-l S_{l-}P^{\prime}_{l-1}(x)) \cr}
\eqno(C.4)$$
By using Eq.(C.3), it is easy to see that
$$\eqalign{
 < m_{s^{\prime}} \mid J_{x} \mid m_s > & = -1/ \sqrt{2}
( <m_{s^{\prime}} \mid \epsilon^{\mu}(+1) J_{\mu} \mid m_s >
- < m_{s^{\prime}} \mid  \epsilon^{\mu}(-1)J_{\mu} \mid m_s >) \cr
 < m_{s^{\prime}} \mid J_{y} \mid m_s > & = i/ \sqrt{2}
( < m_{s^{\prime}} \mid \epsilon^{\mu}(+1) J_{\mu} \mid m_s >
+ < m_{s^{\prime}} \mid \epsilon^{\mu}(-1)J_{\mu} \mid m_s > \cr}
\eqno(C.5a)$$
By using the current conservation condition $k_0 J_0 = \vec{k}\cdot\vec{J}
=k J_z$, Eq.(C.1) for $\lambda=0$ leads to
$$\eqalign{
 < m_{s^{\prime}} \mid J_{z} \mid m_s > & = {\sqrt{-k^2} \over
k_0 }
<m_{s^{\prime}} \mid \epsilon^{\mu}(0)J_{\mu} \mid m_s > \cr}
\eqno(C.5b)$$
The matrix elements defined in Eq.(C.5) are needed to calculated
various polarization observables
defined in Ref.9.

\bigskip
\noindent{\bf APPENDIX D: MODIFICATION OF INVARIANT AMPLITUDES IN THE CASE OF
UNEQUAL CHARGED AND NEUTRAL PION MASSES.}
\medskip
In ref.[7] is was observed that isospin breaking effects due to the difference
of the neutral and charged pion masses turn out to be quite sizeable in the
case of neutral pion photoproduction close to threshold. From this experience
one expects that this isospin breaking effect will also be relevant for
$\pi^0$ electroproduction from protons. Therefore we will display here the
relevant modifications of the invariant amplitudes $A_i(s,u)$ for the reaction
$\gamma^* p \to \pi^0 p$ which arise if we do not set equal the neutral and
charged pion mass as it was done in the chapter IV. Notice that in
contrast to the photoproduction case, one has to consider diagrams from
all three gauge invariant classes.
\medskip
\noindent{\underbar{Class I and II:}}

The expressions given in eqs.(4.5,4.6,4.7) hold still for the invariant
amplitudes. There is only some change in the loop functions $\gamma_i(s,k^2)$
and $\Gamma_i(s,k^2)$ of the following form.  $\gamma_i(s,k^2)$ has to be
evaluated with $M_\pi =
M_{\pi^+}$ and is denoted $\gamma_{i+}(s,k^2)$ whereas $\Gamma_i(s,k^2)$ has to
be calculated with $M_\pi = M_{\pi^0}$ denoted by $\Gamma_{i0}(s,k^2)$.
\medskip
\noindent{\underbar{Electric and magnetic Born terms:}}

Here, the proton Dirac form factor $F_1^p(k^2)$ in eq.(3.12) gets modified.
The $\tilde \gamma_i(k^2) $ and $\tilde \Gamma_i(k^2)$ are replaced by
$\tilde\gamma_{i+}(k^2) $ and $\tilde \Gamma_{i0}(k^2)$, respectively.
Furthermore, $   {\dbar J}_{22}^{\pi\pi}(k^2) $ of eq.(B.1) has to be evaluated
with $M_\pi = M_{\pi^+}$. Of course, now the counter term contribution
proportional to $b^r_9(\lambda) + \tilde b_9^r(\lambda)$ has to be readjusted
in order to reproduce the empirical proton mean square charge radius.
In a similar fashion the value of the magnetic moment counter term $\delta
\kappa_p$ is obtained from
$$\delta \kappa_p = \kappa_p + 8 g_{\pi N}^2 m^2 \bigl\{ \Gamma_{40}(m^2,0)-
2\,   \gamma_{4+}(m^2,0)  \bigr\} = 1.291  \eqno(D.1)$$
\medskip
\noindent{\underbar{Class III:}}

Most of the terms in the $B_i(s,u)$ can be taken over from eq.(4.8) by
simply replacing the respective loop functions by $\gamma_{i+}(s,k^2),\,\,
\Gamma_{i0}(s,k^2),\,\, \tilde \Gamma_{10}(t),\,\, G_{i0}(s,t,k^2)$ and
$\Omega_{i+}(s,u,k^2)$.
In all kinematical prefactors like $M_\pi^2 - t - k^2$,
$M_\pi$ reads $M_{\pi^0}$. The modified loop functions $\Omega_{i+}$ need some
explanation since they involve both the neutral and charged pion.
The
proper denominator for their Feynman parameter representation is now
$$\eqalign{h_{\Omega +}(x,y,z;s,u,k^2) = & M_{\pi^+}^2(1-y) +
M^2_{\pi^0}x(x-1)y^2 +m^2 y^2 +
(s-m^2) (1-x)y (y-1)z \cr & + (u-m^2) x y(1-y)(z-1) + k^2 (1-y)^2 z(z-1)\,.
\cr }  \eqno(D.2)$$
The amplitudes $B_{1,2,3}(s,u)$ involve the combination
$$\Gamma^{\pi N}_0(s) - {M_\pi^2 \over s - m^2 } \underline \Gamma^{\pi N}_0(s)
- \Gamma_2^{\pi N}(s)\,. \eqno(D.3)$$
It gets replaced by
$$2\, \Gamma^{\pi N}_{0+}(s) - {2M_{\pi^+}^2 \over s - m^2 } \underline
{\Gamma}^{\pi N}_{0+}(s) -
2\, \Gamma_{2+}^{\pi N}(s) -\, \Gamma^{\pi N}  _{00}(s) +
{M_{\pi^0}^2 \over s - m^2 } \underline {\Gamma}^{\pi N}_{00}(s) +\,
\Gamma_{20}^{\pi N}(s) \eqno(D.4)$$
where the Feynman parameter representation of the loop functions reads
$$ \Gamma^{\pi N}_{i+}(s) = {1 \over 16 \pi^2} \int_0^1 dx \int_0^1 dy {y
g_i(x,y) \over M_{\pi^+}^2 (1-y) + M_{\pi^0}^2 xy^2(x-1) + m^2 y^2 + (s-m^2)
xy(y-1) } \eqno(D.5)$$
with $g_0 = 1,\, g_1 = 1-y,\, g_2 = 1-xy$ and $\Gamma_{i0}^{\pi N}(s)$ is
obtained if $M_{\pi^+}$ is set equal $M_{\pi^0}$.

\bigskip
\bigskip
\noindent{\bf REFERENCES}
\medskip
\item{1.}V. Bernard, N. Kaiser and Ulf-G. Mei{\ss}ner,
{\it Nucl. Phys.\/}
{\bf B383} (1992) 442.
\smallskip
\item{2.}E. Amaldi, S. Fubini and G. Furlan,
{\it Pion--Electroproduction},
Springer Verlag, Berlin 1979.
\smallskip
\item{3.}T. P. Welch et al., {\it Phys. Rev. Lett.\/}
{\bf 69} (1992) 2761.
\smallskip
\item{4.}Ulf-G. Mei{\ss}ner,
{\it Rep. Prog. Phys.\/} {\bf 56} (1993) 903.
\smallskip
\item{5.}J. Gasser and H. Leutwyler, {\it Ann. Phys. (N.Y.)\/}
 {\bf 158} (1984) 142;

J. Gasser and H. Leutwyler, {\it Nucl. Phys.\/}
 {\bf B250} (1985) 465.
\smallskip
\item{6.}G. Ecker, J. Gasser, A. Pich and E. de Rafael,
{\it Nucl. Phys.\/} {\bf B321}
(1989) 311;

J. F. Donoghue, C. Ramirez and G. Valencia,
{\it Phys. Rev.\/} {\bf D39}
(1989) 1947.
\smallskip
\item{7.}V. Bernard, N. Kaiser and Ulf-G. Mei{\ss}ner,
``Testing nuclear QCD: $\gamma p \to \pi^0 p$ at threshold'',
in the $\pi N$ {\it Newsletter} No. 7 (1992), 62.
\smallskip
\item{8.}S.
   Nozawa and T.-S. H. Lee,  {\it Nucl. Phys.\/} {\bf A513}
(1990) 511.
\smallskip
\item{9.}S.
   Nozawa and T.-S. H. Lee,  {\it Nucl. Phys.\/} {\bf A513}
(1990) 543.
\smallskip
\item{10.}F.A. Berends, A. Donnachie and D.L. Weaver,
{\it Nucl. Phys.\/}
 {\bf B4} (1967) 1.
\smallskip
\item{11.}J. Gasser, M.E. Sainio and A. ${\check {\rm S}}$varc,
{\it Nucl. Phys.\/}
 {\bf
B 307} (1988) 779.
\smallskip
\item{12.}S. R. Amendolia et al., {\it Nucl. Phys.\/} {\bf
B277} (1986) 168.
\smallskip
\item{13.}T. Kitagaki
{\it et al.}, {\it Phys. Rev.\/} {\bf D28}
(1983) 436;

L.A. Ahrens
{\it et al.}, {\it Phys. Rev.\/} {\bf D35}
(1987) 785;

L.A. Ahrens
{\it et al.}, {\it Phys. Lett.\/} {\bf B202}
(1988) 284.
\smallskip
\item{14.}V. Bernard, N. Kaiser and Ulf-G. Mei{\ss}ner,
{\it Phys. Lett.\/} {\bf B282} (1992) 448.
\smallskip
\item{15.}R. Beck {\it et al.}, {\it Phys. Rev. Lett.\/} {\bf 65} (1990) 1841.
\smallskip
\item{16.}S. Fubini, G. Furlan and C. Rossetti,
{\it Nuovo Cim.\/} {\bf 40} (1965) 1171;

Riazuddin and B.W. Lee,
{\it Phys. Rev.\/} {\bf 146} (1966) B1202;

S.L. Adler and F.J. Gilman,
{\it Phys. Rev.\/} {\bf 152}
(1966) B1460;

S.L. Adler,
{\it Ann. Phys. (N.Y.)\/} {\bf 50}
(1968) 189;

N. Dombey and R.J. Read,
{\it Nucl. Phys.\/} {\bf B60} (1973) 65.
\smallskip
\item{17.}S. Scherer and J. H. Koch,
{\it Nucl. Phys.\/} {\bf A534}
(1991) 461.
\smallskip
\item{18.}Y. Nambu and D. Luri\'e, {\it Phys. Rev.\/} {\bf 125}
(1962) 1429;

Y. Nambu and E. Shrauner, {\it Phys. Rev.\/} {\bf 128}
(1962) 862.
\smallskip
\item{19.}V. Bernard, N. Kaiser and Ulf-G. Mei{\ss}ner,
{\it Phys. Rev. Lett.\/} {\bf 69} (1992) 1877.
\smallskip
\item{20.}S. Weinberg,
{\it Physica\/} {\bf 96A}
(1979) 327.
\smallskip
\item{21.}I.A. Vainshtein and V.I. Zakharov,
 {\it Nucl. Phys.\/} {\bf B36} (1972) 589.
\smallskip
\item{22.}P. de Baenst,
{\it Nucl. Phys.\/} {\bf B24} (1970) 633.
\smallskip
\item{23.}S. Scherer, J. H. Koch and J. L. Friar,
{\it Nucl. Phys.\/} {\bf A552} (1993) 515.
\smallskip
\item{24.}K. Ohta,
{\it Phys. Rev.\/} {\bf C47} (1993) 2344.
\smallskip
\item{25.}N.M. Kroll and M.A. Ruderman,
{\it Phys. Rev.\/} {\bf 93} (1954) 233.
\smallskip
\item{26.}E. Jenkins and A.V. Manohar, {\it Phys. Lett.\/} {\bf B255} (1991)
558.
\smallskip
\item{27.}V. Bernard, N. Kaiser, J. Kambor
and Ulf-G. Mei{\ss}ner, {\it Nucl. Phys.\/} {\bf B388} (1992) 315.
\smallskip
\item{28.}J. Gasser and Ulf-G. Mei{\ss}ner,
{\it Nucl. Phys.\/} {\bf B357} (1991) 90.
\smallskip
\item{29.}J. Gasser and H. Leutwyler,
{\it Phys. Reports\/} {\bf C87} (1982) 77.
\smallskip
\item{30.}V. Bernard, N. Kaiser, T.--S. H. Lee and Ulf-G. Mei{\ss}ner,
{\it Phys. Rev. Lett.\/} {\bf 70} (1993) 367.
\smallskip
\item{31.}A. del Guerra  {\it et al.}, {\it Nucl. Phys.\/}
 {\bf B107} (1976) 65;

M.G. Olsson, E.T. Osypowski and E.H. Monsay, {\it
Phys. Rev.\/} {\bf D17} (1978) 2938.
\smallskip

\vfill \eject
\medskip \nobreak
\centerline{\bf FIGURE CAPTIONS}
\bigskip
\item{Fig.1.} The functions $\Xi_{1,2,3,4} (\rho )$ defined in eq.(5.2) with
$\rho = -k^2 / M_\pi^2$. The solid, dotted, dashed and dash--dotted lines
refer to $\Xi_1$, $\Xi_2$, $\Xi_3$ and $\Xi_4$, respectively.
\medskip
\item{Fig.2.} The functions $\Phi_{1,2} (k^2 )$ defined in eq.(5.5).
The solid and dotted lines refer to $\Phi_1$ and $\Phi_2$, in order.
\medskip
\item{Fig.3.} The functions $\tilde{\Xi}_{1,2} (\rho_0 )$ defined in eq.(5.9)
with $\rho_0 = -k^2 / M_{\pi^0}^2$. The solid and dotted
lines refer to $\tilde{\Xi}_1$ and $\tilde{\Xi}_2$, respectively.
\medskip
\item{Fig.4.} Comparison of the theoretical predictions with the recent NIKHEF
data for $p(\gamma^\star ,\pi^0 )p$ [3]. The
solid circles, open circles and open squares
refer to the one--loop CHPT, tree and PV Born term calculation, in order.
\medskip
\item{Fig.5} The longitudinal $S_{0+}$ multipole amplitudes for
    $p(\gamma^\star ,\pi^+ )n$,     $n(\gamma^\star ,\pi^- )p$ and
$p(\gamma^\star ,\pi^0 )p$.
The solid, dotted and dashed lines refer to the one--loop CHPT, the tree and
the PV Born calculation, in order.
\medskip
\item{Fig.6} The transversal $E_{0+}$ multipole amplitudes for
    $p(\gamma^\star ,\pi^+ )n$,     $n(\gamma^\star ,\pi^- )p$ and
$p(\gamma^\star ,\pi^0 )p$.
\medskip
\item{Fig.7} $k^2$--dependence of $\sigma_T$ and $\sigma_L$ for
$p(\gamma^\star ,\pi^+ )n$,
    $ n(\gamma^\star ,\pi^-)p$ and     $p(\gamma^\star , \pi^0 )p$.
\medskip
\item{Fig.8} Angular distributions of the transverse cross sections for
    $p(\gamma^\star ,\pi^+)n$,
    $ n(\gamma^\star ,\pi^-)p$ and     $p(\gamma^\star ,\pi^0 )p$.
For notations see Fig.5.
\medskip
\item{Fig.9} Same as Fig.8 except for the longitudinal cross sections.
\medskip
\item{Fig.10} Same as Fig.8 except for the interference cross sections.
\medskip
\item{Fig.11} Same as Fig.8 except for the polarization cross sections.
\medskip
\item{Fig.12} Dependence of the axial cut--off mass for
$p(\gamma^\star ,\pi^+)n$.    We show
the angular distributions     $d\sigma_T / d\Omega$, $d\sigma_L / d\Omega$,
    $d\sigma_I / d\Omega$ and $d\sigma_I / d\Omega$ at $k^2 = -0.04$
GeV$^2$ for $M_A= 0.96, 1.06, 1.16$ GeV corresponding to the dotted, solid and
dashed lines, respectively.
\medskip
\item{Fig.13} The S--wave cross section $a_0$ for
$p(\gamma^\star ,\pi^0 )p$ including isospin--breaking for $W=1074$ MeV
and $\epsilon = 0.58$. The solid,
dotted and dashed lines represent the isosymmetric CHPT, CHPT with
isospin--breaking and PV predictions, in order.
\medskip
\item{Fig.14} Angular distributions for
$p(\gamma^\star ,\pi^0 )p$ including isospin--breaking.
For notations see Fig.13.
\medskip
\item{Fig.15} $k^2$--dependence of $\sigma_T$ and $\sigma_L$ for
$p(\gamma^\star ,\pi^0 )p$ including isospin--breaking.
For notations see Fig.13.
\vfill \eject
\smallskip \nobreak
\centerline{\bf TABLES}
\bigskip
$$\hbox{\vbox{\offinterlineskip
\def\strut{\hbox{\vrule height 10pt depth 10pt width 0pt}}
\hrule
\halign{
\strut\vrule# \tabskip 0.1in &
\hfil#\hfil  &
\vrule# &
\hfil#\hfil &
\hfil#\hfil &
\hfil#\hfil &
\vrule# \tabskip 0.0in
\cr
\noalign{\hrule}
& $k^2$
&& --0.001 GeV$^2$  & --0.04 GeV$^2$  & --0.06 GeV$^2$
& \cr
\noalign{\hrule}
&   $E_{0+}$
&&( 25.77,  0.2663) &( 15.00,  0.5830) & ( 9.899,  0.6602) & \cr
&   $S_{0+}$
&&(-18.44, 0.1890) &( -21.32, 0.2296) & (-22.58, 0.2425) & \cr
&   $E_{1+}$
&&(0.7625, 2.022 10$^{-6}$) &(0.6366, -1.169 10$^{-6}$)
& (0.5707, -7.883 10$^{-8}$) & \cr
&   $S_{1+}$
&&(-0.4098, 2.631 10$^{-6}$) &(-0.3335, 7.441 10$^{-6}$)
& (-0.2849, 9.871 10$^{-6}$) & \cr
&   $S_{1-}$
&&(-3.373, 9.221 10$^{-3}$) &(-4.831, 0.01405) & (-5.152, 0.01523) & \cr
&   $M_{1+}$
&&(-1.037, -5.792 10$^{-5}$) &(-1.135, -9.365 10$^{-5}$)
& (-1.168, -1.058 10$^{-4}$) & \cr
&   $M_{1-}$
&&(0.2250, -6.891 10$^{-5}$) &(-0.8534, 2.085 10$^{-3}$)
& (-1.271, 2.989 10$^{-3}$) & \cr
\noalign{\hrule}}}}$$
\smallskip
{\noindent\narrower Table 1a:\quad Various multipoles for the reaction
$\gamma^\star p \to \pi^+ n$ in units of $10^{-3}/M_{\pi^+}$.
\bigskip}
\medskip
$$\hbox{\vbox{\offinterlineskip
\def\strut{\hbox{\vrule height 10pt depth 10pt width 0pt}}
\hrule
\halign{
\strut\vrule# \tabskip 0.1in &
\hfil#\hfil  &
\vrule# &
\hfil#\hfil &
\hfil#\hfil &
\hfil#\hfil &
\vrule# \tabskip 0.0in
\cr
\noalign{\hrule}
& $k^2$
&& --0.001 GeV$^2$  & --0.04 GeV$^2$  & --0.06 GeV$^2$
& \cr
\noalign{\hrule}
&   $E_{0+}$
&&(-29.99, -0.3543) &(-16.96, -0.6121) & (-10.86, -0.6657) & \cr
&   $S_{0+}$
&&(22.98, -0.08447) &(26.55, -0.08553) & (27.24, -0.08384) & \cr
&   $E_{1+}$
&&(-0.7534, -4.067 10$^{-11}$) &(-0.6263, 3.178 10$^{-9}$)
& (-0.5611, 5.027 10$^{-9}$) & \cr
&   $S_{1+}$
&&(0.4010, -1.952 10$^{-10}$) &(0.3174, -7.329 10$^{-10}$)
& (0.2691, -1.377 10$^{-9}$) & \cr
&   $S_{1-}$
&&(3.336, -8.639 10$^{-3}$) &(4.778, -0.01237) & (5.137, -0.01300) & \cr
&   $M_{1+}$
&&(1.242, 3.158 10$^{-11}$) &(1.350, -3.199 10$^{-9}$)
& (1.341, -5.050 10$^{-9}$) & \cr
&   $M_{1-}$
&&(-0.5396, 1.667 10$^{-3}$) &(0.5531, 3.483 10$^{-4}$)
& (1.059, -1.967 10$^{-4}$) & \cr
\noalign{\hrule}}}}$$
\smallskip
{\noindent\narrower Table 1b:\quad Various multipoles for the reaction
$\gamma^\star n \to \pi^- p$ in units of $10^{-3}/M_{\pi^+}$.
\medskip}
$$\hbox{\vbox{\offinterlineskip
\def\strut{\hbox{\vrule height 10pt depth 10pt width 0pt}}
\hrule
\halign{
\strut\vrule# \tabskip 0.1in &
\hfil#\hfil  &
\vrule# &
\hfil#\hfil &
\hfil#\hfil &
\hfil#\hfil &
\vrule# \tabskip 0.0in
\cr
\noalign{\hrule}
& $k^2$
&& --0.001 GeV$^2$  & --0.04 GeV$^2$  & --0.06 GeV$^2$
& \cr
\noalign{\hrule}
&   $E_{0+}$
&&(-1.304, 0.5103) &(1.625, 0.9461) & (2.668, 1.045) & \cr
&   $S_{0+}$
&&(3.199, 0.1805) &(3.758, 0.2102) & (3.386, 0.2176) & \cr
&   $E_{1+}$
&&(6.989 10$^{-3}$, -2.634 10$^{-4}$) &(1.609 10$^{-3}$, -4.267 10$^{-4}$)
& (-2.648 10$^{-3}$, -4.385 10$^{-4}$) & \cr
&   $S_{1+}$
&&(-0.01311, -7.260 10$^{-5}$) &(-0.03610, -7.342 10$^{-5}$)
& (-0.04627, -7.053 10$^{-5}$) & \cr
&   $S_{1-}$
&&(0.09471, 0.01126) &(0.2800, 0.01760) & (0.3872, 0.01912) & \cr
&   $M_{1+}$
&&(0.1162, 2.884 10$^{-4}$) &(0.04723, 9.844 10$^{-5}$)
& (-0.02483, 6.400 10$^{-5}$) & \cr
&   $M_{1-}$
&&(-1.521, -1.233 10$^{-4}$) &(-2.303, 2.877 10$^{-3}$)
& (-2.520, 4.016 10$^{-3}$) & \cr
\noalign{\hrule}}}}$$
\smallskip
{\noindent\narrower Table 1c:\quad Various multipoles for the reaction
$\gamma^\star p \to \pi^0 p$ in units of $10^{-3}/M_{\pi^+}$.
\bigskip}
\bigskip
$$\hbox{\vbox{\offinterlineskip
\def\strut{\hbox{\vrule height 12pt depth 12pt width 0pt}}
\hrule
\halign{
\strut\vrule# \tabskip 0.1in &
\hfil#\hfil  &
\vrule# &
\hfil#\hfil &
\hfil#\hfil &
\hfil#\hfil &
\vrule# \tabskip 0.0in
\cr
\noalign{\hrule}
&  $M_A [{\rm GeV}]$
&& $1.06$  & $1.16$   & $0.96$
& \cr
\noalign{\hrule}
& Re$E_{0+} [10^{-3}/M_{\pi^+}]$
&&  15.00         &  15.98        & 13.71 & \cr
& Re$S_{0+} [10^{-3}/M_{\pi^+}]$
&& -21.32          &  -20.42        & -22.51 & \cr
& $\sigma_T [\mu {\rm b}]$
&& 10.81            & 12.17       & 9.04  & \cr
&  $\sigma_L [\mu {\rm b}]$
&& 16.87            & 15.51         & 18.70  & \cr
&  $(\sigma_T + \sigma_L)
[\mu {\rm b}]$
&& 27.68            & 27.68         & 27.68  & \cr
& $\sigma_T / \sigma_L $
&& 0.64            & 0.79         & 0.48  & \cr
\noalign{\hrule}}}}$$
\smallskip
{\noindent\narrower Table 2a:\quad Dependence on the axial cut--off
mass $M_A$ of  various observables for the reaction
$\gamma^\star p \to \pi^+ n$.
\bigskip}
$$\hbox{\vbox{\offinterlineskip
\def\strut{\hbox{\vrule height 12pt depth 12pt width 0pt}}
\hrule
\halign{
\strut\vrule# \tabskip 0.1in &
\hfil#\hfil  &
\vrule# &
\hfil#\hfil &
\hfil#\hfil &
\hfil#\hfil &
\vrule# \tabskip 0.0in
\cr
\noalign{\hrule}
&  $M_A [{\rm GeV}]$
&& $1.06$  & $1.16$   & $0.96$
& \cr
\noalign{\hrule}
& Re$E_{0+} [10^{-3}/M_{\pi^+}]$
&& -16.96         & -17.94        & -15.67 & \cr
& Re$S_{0+} [10^{-3}/M_{\pi^+}]$
&& 26.55          &  25.65        & 27.74 & \cr
&  $\sigma_T [\mu {\rm b}]$
&& 13.70            & 15.27         & 11.74  & \cr
&  $\sigma_L [\mu {\rm b}]$
&& 25.58            & 23.95         & 27.88  & \cr
&  $(\sigma_T + \sigma_L)
[\mu {\rm b}]$
&& 39.28            & 39.21         & 39.62  & \cr
& $\sigma_T / \sigma_L $
&& 0.54            & 0.64         & 0.42  & \cr
\noalign{\hrule}}}}$$
\smallskip
{\noindent\narrower Table 2b:\quad Dependence on the axial cut--off
mass $M_A$ of  various observables for the reaction
$\gamma^\star n \to \pi^- p$.
\vfill
\eject

\end